\documentclass[a4paper]{scrartcl}
\usepackage[normalem]{ulem}
\usepackage[utf8]{inputenc}
\usepackage{hyperref}
\usepackage{float}
\usepackage[table,xcdraw]{xcolor}
\usepackage{amsmath}
\usepackage{color}
\usepackage{setspace}
\usepackage{xr}
\usepackage{graphicx}
\usepackage{amsfonts}
\usepackage{amsmath}
\usepackage{amssymb}
\usepackage{mathtools}
\usepackage{cite}
\usepackage{xspace}
\usepackage{float}
\usepackage{subcaption}
\usepackage{color}
\usepackage{tikz}
\usepackage{booktabs}
\usepackage{apacite}
\usepackage{makecell}
\usepackage{comment}
\usepackage{bm}
\usepackage{svg}
\usepackage{algorithm}
\usepackage{algpseudocode}
\usepackage[T1]{fontenc}
\usepackage[title]{appendix}
\usepackage{times}
\usepackage{tikz}

\newlength{\insidefboxwidth}
\newcommand\mymid{\ {\big|}\ }

\newcolumntype{L}[1]{>{\raggedright\arraybackslash}p{#1}}
\newcolumntype{C}[1]{>{\centering\arraybackslash}p{#1}}
\newcommand{\red}[1]{\text{\textcolor{red}{#1}}}
\newcommand{\blue}[1]{\text{\textcolor{blue}{#1}}}
\newcommand\bp{\text{\textbullet}}

\newcommand\A{\mathcal{A}}
\newcommand\B{\mathcal{B}}

\newcommand\E{\mathcal{E}}

\newcommand\Hs{\mathcal{H}}
\newcommand\K{\mathcal{K}}
\newcommand\Lab{\mathcal{L}}
\newcommand\N{\mathcal{N}}
\newcommand\R{\mathcal{R}}

\newcommand\T{\mathcal{T}}

\newcommand\V{\mathcal{V}}

\newcommand\Y{\mathcal{Y}}
\newcommand\Z{\mathcal{Z}}
\newcommand\nv{\blue{nil}}

\newcommand\sem[1]{{[\![ #1 ]\!]}}

\begin{document}
\setcounter{secnumdepth}{5}
\title{The analysis approach of ThreatGet}
\subtitle{(version 21.04)}
\author{Korbinian Christl\\
    \href{mailto:Korbinian.Christl@ait.ac.at}{Korbinian.Christl@ait.ac.at} \\ 
    Thorsten Tarrach\\
    \href{mailto:Thorsten.Tarrach@ait.ac.at}{Thorsten.Tarrach@ait.ac.at}}
\publishers{AIT Austrian Institute of Technology}
%
%

\maketitle

\begin{abstract}
Nowadays, almost all electronic devices include a communication interface that allows to interact with them, exchange data, or operate their services remotely. The trend toward increased interconnectivity simultaneously increases the vulnerability of these systems. Due to the high costs associated with comprehensive security analysis, many manufacturers neglect the safety aspect of a product in order to avoid costs. However, the importance of secure IT systems is growing, as the security of a system can also influence safety-critical aspects. Standard security analysis approaches are nowadays still mainly based on time-intensive and error-prone manual activities. In this paper, we present the formal concepts of the automatic threat and vulnerability analysis tool ThreatGet. Therefore, we introduce the concept of the Extended Data-Flow Diagram that is used to represent the system under investigation in an abstracted form, and we highlight the formal analysis language of the tool. This domain-specific language is used to formulate so-called anti-patterns. These anti-patterns that can be interpreted by the tool for an automatic security analysis of the system. Besides the language declaration, we present the entire semantic evaluation of the language during the analysis. Parts of the definitions and elaborations of the diagram model and the analysis language were developed in the context of the master thesis of Korbinian Christl, in cooperation with the University of Vienna. 
\end{abstract}

\clearpage

\tableofcontents

\clearpage

\section{Introduction}
\label{sec:intro}

Over the past two decades, IT-systems, their areas of application and their purposes have significantly evolved. As a result, these systems have become increasingly complex and now offer cross-domain interfaces to exchange data and services. \\
This development has meant that everyday devices such as coffee machines and entire production lines in factories can be controlled and monitored from remote. Moreover, this development process is far from complete, as the efforts toward the autonomous driving show.  \\

Besides the positive aspects that this development has brought, negative factors have also emerged. Due to the rapid development of new products and the competition on the market, manufacturers often omit security and safety audits of their products to save money \shortcite{gilchrist_iot_2017}.  However, as the number of electronic devices and their communication interfaces grows, the vulnerability of the systems increases  \shortcite{jang_jaccard_survey_2014}. Attacks by the "Mirai" botnet and the rise of malicious ransomware show the impact that insecure electronic devices can have \shortcite{mos_growing_2020}.
As the automotive industry moves toward autonomous driving, the importance of vehicle security becomes even more apparent, as security deficiencies in this area can compromise passenger safety. This concern has been impressively demonstrated by \shortciteA{miller_remote_2015} in their paper on the "Jeep Cherokee" hack \shortcite{miller_remote_2015}.\\
With increasing customer awareness for security issues and rising interest in data privacy and security, manufacturers face the challenge of making their products not only smarter, but also safer and more secure. \\

The central issue in developing safe and secure products is that most security analysis methods are still predominantly based on error-prone and time-consuming manual activities. Therefore, high costs may be incurred, especially if an analysis has to be repeated due to changes to the system. Moreover, the quality of the results of such an analysis depends heavily on the knowledge and experience of the analyst \shortcite{yilmaz_assessing_2020}. \\
Another challenge stems from the fact that publicly available threat and vulnerability databases such as the "National Vulnerability Database" (NVD), "Common Vulnerabilities and Exposures" (CVE), "Common Vulnerabilities Enumeration" (CWE), and the "Common Attack Pattern Enumeration and Classification" (CAPEC) are enormous collections of entries that are almost daily updated, and no analyst can keep track of all the new entries \shortcite{nvd_nvd_2020, mitre_cve_2019, mitre_capec_2020, mitre_cwe_2020}. \\
Finally we want to highlight the fact that compliance with security and safety standards such as the ISO/SAE 21434 \shortcite{iso_21434} will become mandatory in the future. As a result, smart device manufacturers will be required to ensure the security of their products.  However, there are probably not enough security experts to cover all areas and domains \shortcite{schmittner_preliminary_2020}.\\
These challenges and circumstances indicate a high demand for an automated, repeatable, and objective solution to identify vulnerabilities and potential threats within a system. In addition, it should be applicable already during the planning- as well as the design-phase of a project to avoid the costs of an retrospective solution \shortcite{shostack_threat_2014}. \\

In order to tackle these challenges, the Dependable Systems Engineering Group (DSE) at the Austrian Institute of Technology (AIT) has developed and maintains the ThreatGet tool. Furthermore, this tool is already commercially distributed in cooperation with LieberLieber and is also available for academic purposes on a test basis. \\
The concept of ThreatGet is based on two fundamental components comparable to other software/\allowbreak system-oriented analysis tools. \autoref{sec:related} of discusses these tools and highlights further related research in this area. \\
On the one hand, the first component is called the \emph{system-model}. It is a graphical representation of the system under study in an abstracted diagram. ThreatGet uses an extended form of the data-flow diagram (DFD), which is also discussed in this paper. Within this diagram, the user defines the structure of the system as well as all security and safety-related aspects \shortcite{hussain_threat_2014,shostack_threat_2014}. \\
On the other hand, the second component is often referred to as the \emph{threat-model}. The term threat-model can be used to describe any type of knowledge collection that focuses on the definition and retention of threats or vulnerabilities. ThreatGet uses a proprietary analysis language that allows users to define threats in a domain-specific context that both humans and machines can interpret. In this way, ThreatGet provides a viable approach to managing threat intelligence and a sustainable approach to reusing the gathered knowledge. \\
In section 3, both the concept of the system-model and the threat model are presented and explained in detail. In addition to the formal definitions of the diagram and the language, their application is also explained using examples. Based on the language specification, it is explained in detail how its semantic evaluation is conducted based on the definition of the "Advanced Data-Flow Diagram" (ADFD). These specifications allow users to precisely understand how the expressions formulated in the analysis language are processed, how the analysis itself is conducted, and how the results are obtained. \\

Parts of this paper, especially \autoref{sec:analysis_approach}, were written in cooperation with the University of Vienna as part of the master thesis of Korbinian Christl. The content of this chapter has been largely adopted but adjusted accordingly to the modifications of the analysis language.\\
This paper is a direct extension of the public tool documentation available on the public homepage\footnote{\url{https://www.threatget.com/}}. Furthermore, it is considered a living document, which will be updated accordingly to the modifications of the analysis language or the diagram model used.  
\section{The Analysis Approach}
\label{sec:analysis_approach}
The analysis approach of ThreatGet includes three components: an "Advanced Data-Flow Diagram" (ADFD) that serves as a system-model, a rule-based threat-model, and an analysis engine that compares the system-model with the threat-model. This section is divided into three subsections. \\

The first section outlines several preliminaries and notational conventions used in the following sections for the formal definition of the system-model and the threat-model. \\

The second section illustrates the formal definition of the system-model and threat-model. Firstly, the diagram components and the structure of a diagram are presented using a simplified example. This example is used continuously to illustrate the following formal concepts. Building on this, the formal representation of a diagram is presented. In addition, a content-specification is outlined. The content-specification includes all predefined stencils. Therefore, it can be used to validate the content of the defined diagram as well as the anti-patterns. It serves as a bridge between the system-model and the threat-model.    \\

Afterward, the formal definition of the rule-based threat-model using the analysis language is elaborated. Each rule describes a potential threat in the form of an anti-pattern. Threats are defined and stored in a specially defined language, whose syntax and semantics are explained in detail.

\subsection{Preliminaries and Notational Conventions}
\label{sec:preliminaries}
Several formal mathematical definitions are presented throughout the following sections. These definitions use the preliminaries and notational conventions presented in this section.

\begin{itemize}
    \setlength\itemsep{0.3cm}
     \setlength\itemsep{0.3cm}
    \item A relation $R$ over a set $A$ is \emph{transitive} if $\forall a,b,c \in A$,  $(a,b) \in R \wedge (b,c) \in R \implies (a,c) \in R $ \cite{weisstein_transitive_2021}.
    
    \item A relation $R$ over a set $A$ is \emph{irreflexive} if $\forall a \in A, (a,a) \notin R$
    \cite{weisstein_irreflexive_2021}.
    
    \item A relation $R$ over a set $A$ is \emph{antisymmetric} if $\forall a,b \in A, (a,b)\in R \implies (b,a)\notin R$ \cite{weisstein_antisym_2021}.

    \item A function $f$ is a \emph{total} function if and only if it maps each value of the domain $(x \in X)$ to a value of the co-domain $Y$, denoted as $f: X \rightarrow Y$ \cite{black_totalfunction_2007}.
    
    \item A function $f: X \rightarrow Y$ is a  \emph{partial} function if $f$ maps only values from a proper subset $S$ of $X$ to $Y$, denoted as $f :X \nrightarrow Y $ \cite{black_partialfunction_2019}.
    
    \item The result of a partial function $f:X \nrightarrow Y$ is \emph{undefined} for a parameter $x \in X$ if the function does not provide any value  $y \in Y$, such that $f(x) = \emph{undefined}$  .
    
    \item The result of a partial function $f:X \nrightarrow Y$ is \emph{defined} for a parameter $x \in X$ if the function does provide a value  $y \in Y$, such that $f(x) = y$.
    
    \item An empty function $f$ for a domain $X$ always returns \emph{undefined} for any parameter $x \in X$. $f(x)=\emph{undefined}$.
    
    \item A sequence $seq$ is defined as an ordered set of entries $e$ where the position of each entry is indicated by a number $seq(e_1 \ldots e_i)$
    
    \item The \emph{powerset} $\mathfrak{P}^A$  is defined as a set including subsets of the a set $A$ \cite{weisstein_powerset_2021}.
    
    \item Let f be a partial function. By $f[a:=b]$ it is denoted that the function returns the value $b$ for argument $a$ and $f(x)$ for all other arguments $x$.

\end{itemize}

\autoref{tab:notation} contains the most important notational conventions used throughout the following sections. Generally, sets are labeled with uppercase letters, and their contained entries are labeled with lowercase letters. \\
The exact definition of the individual table entries is explained in the corresponding sections. \\

\begin{table}[htb]
  \centering
  \begin{tabular}{|L{0.3\textwidth} | C{0.3\textwidth}| C{0.3\textwidth}|}
    \toprule
    \textbf{Concept} & \textbf{Instance notation} & \textbf{Set notation}\\
    \midrule
    Element identifiers     & $n$           & $\N$\\
    Element types           & $l$           & $\Lab$\\
    Asset identifiers       & $y$           & $\Y$\\
    Asset types             & $z$           & $\Z$\\
    Boundary identifiers    & $a$           & $\A$\\
    Boundary types          & $b$           & $\B$\\
    Connector identifiers   & $r$           & $\R$\\
    Connector types         & $t$           & $\T$\\
    Property keys           & $k$           & $\K$\\
    Property values         & $v$           & $\V$\\
    Flow                    & $p$           & -\\
    Sequence                & $seq(\ldots)$ & -\\
    Functions               & $f,f_1$        & -\\
    \bottomrule
  \end{tabular}
  \caption{Notational conventions}
  \label{tab:notation}
\end{table}

\subsection{Model Concepts}
\label{sec:diagram_model}
This section presents the formal concepts of the system-model and the threat-model. Firstly, the "Advanced Data-Flow Diagram" (ADFD) and its components are discussed and illustrated using an example. Secondly, the content and structure of the diagram are formally defined and discussed.
Thirdly, the definition of the rule-based threat-model and its language syntax is explained.

\subsubsection{Diagram Components and Modeling Notation}
\label{sec:model_notation}
A standard data-flow diagram does not offer enough semantic depth to meet the requirements of a comprehensive security analysis. This section presents an advanced version of the data-flow diagram (ADFD). Therefore, the individual diagram components, their modeling conventions, and their intended use is described in the following listing:

\begin{itemize}
    \setlength\itemsep{0.3cm}
    
    \item \textbf{Element:} A standard data-flow diagram uses different diagram components to differentiate between external actors, processes, and data stores. However, the ADFD defines a single diagram component called "Element" with an assignable type. A type is predefined as a stencil (template) and describes the nature of an element, and contains its properties. This type-based approach allows both physical and logical system elements to be modeled and can be expanded as required. In addition, elements can be nested within each other. This hierarchical structure allows an additional semantic relationship besides the connectors (data-flows).
    
    \item \textbf{Connector:} A connector indicates an interaction between two diagram elements. A connector can be used to represent logical as well as physical data-flows. Connectors are always directed between a source and a target element. Similar to the elements, connectors are further specified by their assigned type.
     
     \item \textbf{Asset:} Assets describe logical or physical objects of value. An asset can be used to represent system-critical software or the movement of confidential data in a system. Assets are further specified by a predefined type. 

     \item \textbf{Asset Connector:} In contrast to the previous connector, the asset connector is not used to display the relationships between two elements but rather to display the relationship between an asset and an element or connector. An asset connector is undirected, and an asset can be linked to several different elements and connectors. It is the only diagram component that cannot have an assigned type as it only represents the affiliation of an asset to an element or connector.
       
    \item \textbf{Boundary:} A boundary describes a separation between logically, physically, or legally separated system elements. A boundary has an assigned type, but a has no properties. Boundaries are located at the lowest hierarchical level of a diagram and can only be contained by other boundaries.
\end{itemize}

\begin{figure}[htb]
    \centering
    \includegraphics[width=1.0\textwidth]{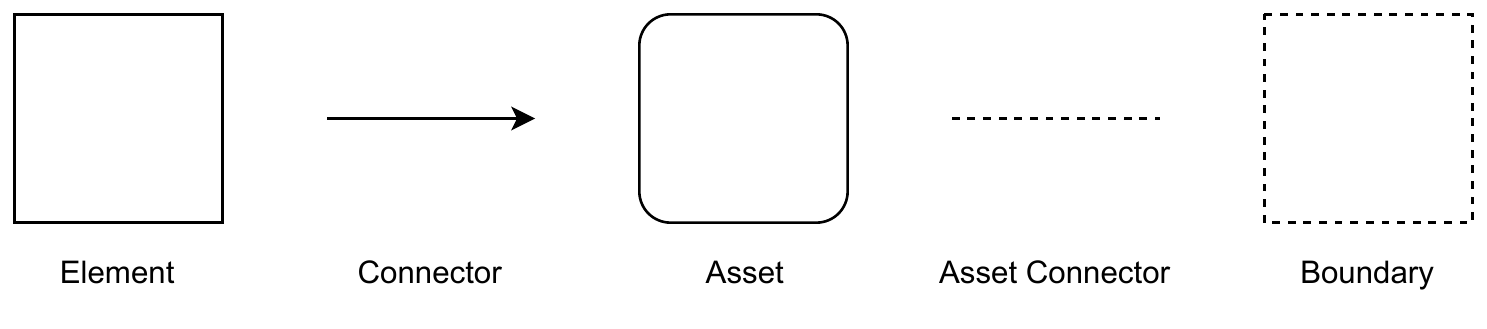}
    \caption{Graphical notation of the individual diagram components}
    \label{fig:diagram_components}
\end{figure}

Properties are also predefined as stencils in the form of key-value pairs. Each property stencil consists of a unique key and multiple values where each value must be unique for the key.  Boundary types have no assigned properties as they are not a real part of the system.\\
\autoref{fig:diagram_components} illustrates the graphical notation of the individual components.

\clearpage

\subsubsection{Diagram Example}
\label{sec:diagram_example}
\begin{figure}[htb]
    \centering
    \includegraphics[width=1.0\textwidth]{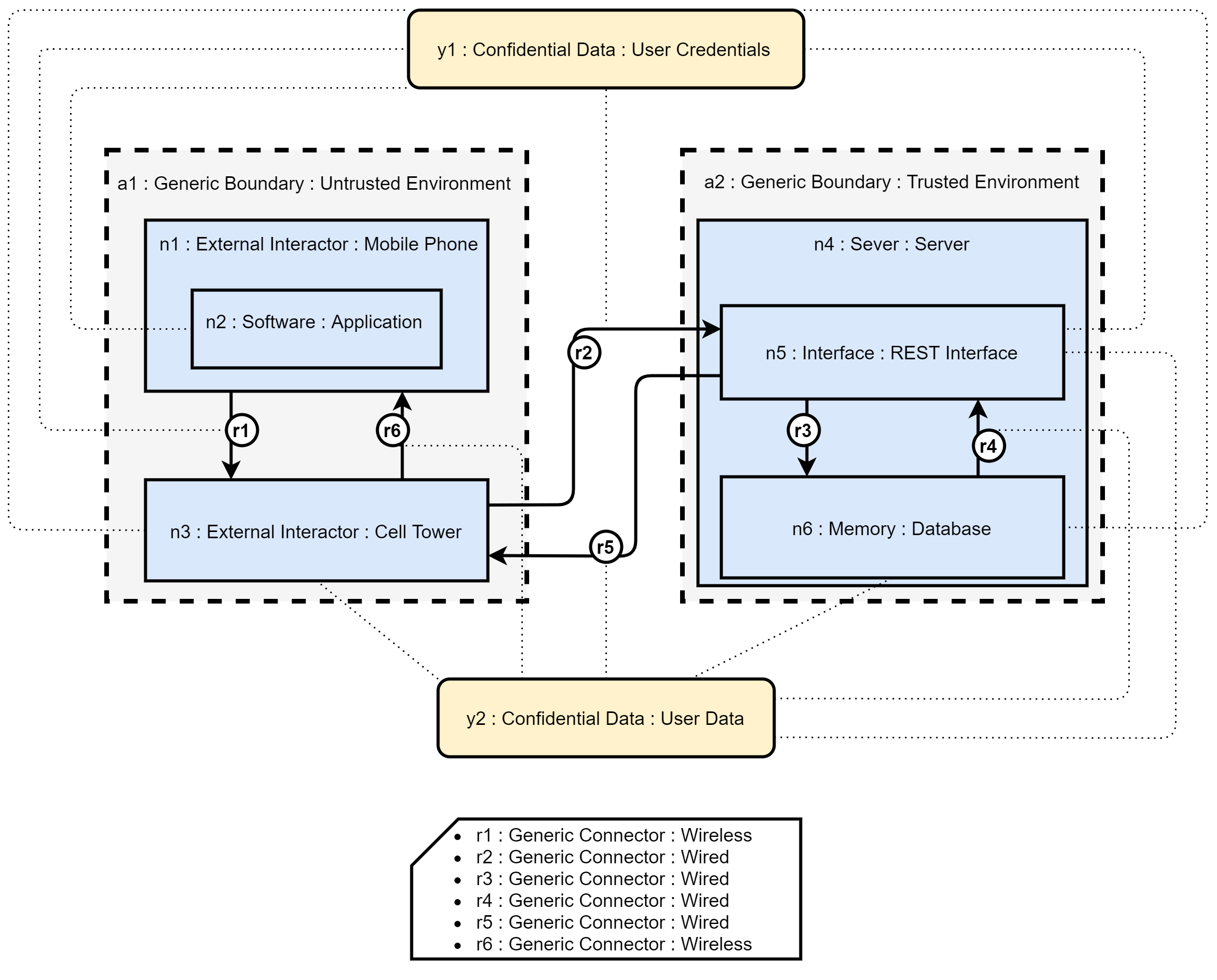}
    \caption{Mobile data request example diagram}
    \label{fig:mobile_phone_diagram}
\end{figure}

The previous section introduced the individual diagram components. This section demonstrates their intended use with a fictitious simplified example. The diagram shown in \autoref{fig:mobile_phone_diagram} describes the request of confidential user data by a mobile phone user. 

The diagram describes the following sequence of events: A user requests data from an offered service by entering the credentials into the mobile device application. The credentials are sent to the nearest cell tower via a wireless connection. From this point, the data is forwarded to the specific server, which in turn validates the provided data and, if it is correct, sends the requested data back the same way. \\
All elements, connectors, boundaries, and assets are annotated according to the same scheme. This scheme has the following structure : \\
\textless Identifier\textgreater: \textless Top-Type\textgreater \textless Sub-Type\textgreater. \\

Each diagram component has a unique identifier, which is used to distinguish the individual components in addition to the assigned types. Identifier and types are separated by a colon. The "Top-Type" represents a generic group of more specific "Sub-Types". For example: Both assets within the diagram are of the "Top-Type: Confidential Data". However, the user enter the "User Credentials" and receives the "User Data". Both are "Sub-Types" of the "Top-Type: "Confidential Data".  \\
The left side of the diagram shows the "Mobile Phone" (n1), the used "Application" (n2), and the closest "Cell Tower" (n3). These three elements are independent of the service offered on the right-hand side and are therefore in an "Untrusted Environment" boundary (a1).  As described above, the user must first provide her login data. The credential information is shown within the diagram as an asset called "User Credentials" (y1). The asset component is linked to the connectors that transport the data asset and the elements that process the data. The link is visualized using asset connectors. The requested service is shown on the right side of the diagram. It consists of a "Server" (n4) which contains a "REST Interface" (n5) and a "Database" (n6). The "REST Interface" receives the user credentials, queries the requested data from the database, and sends it back. This data is also modeled as an asset named "User Data" (y2). \\

\begin{table}[htb]
  \centering
  \begin{tabular}{|L{2cm} |L{3cm} |L{3cm} |L{3cm}|}
    \toprule
    \textbf{Identifier}  & \textbf{Type}  & \textbf{Property Key}  & \textbf{Value} \\
    \midrule
    n1	& Mobile Phone	& OS					&Android 		\\
    n1	& Mobile Phone	& Vendor				&Third Party 		\\
    n2	& Application	& Vendor				&Third Party 		\\
    n4	& Server		& Vendor				&Third Party 		\\
    n5	& REST Interface& Input Validation	&Yes 		\\
    n5	& REST Interface& Input Sanitization	&No 		\\
    n6	& Database		& Encrypted			& Yes 		\\
    r1	& Wireless		& Protocol 			& HTTP 		\\
    r2	& Wired			& Protocol 			& HTTP 		\\
    r3	& Wired			& Protocol 			& HTTP 		\\
    r4	& Wired			& Protocol 			& HTTP 		\\
    r5	& Wired			& Protocol 			& HTTP 		\\
    r6	& Wireless		& Protocol 			& HTTP 		\\
    y1	& User Credentials	& Encrypted		& No 		\\
    y2	& User Data	&Encrypted	&Yes 		\\
    \bottomrule
  \end{tabular}
  \caption{Example Diagram assigned Properties}
  \label{tab:assigned_property_values}
\end{table}

\autoref{tab:assigned_property_values} displays the assigned properties of the individual diagram components. For example, the mobile phone operating system is Android, and the communication protocol of the connectors is HTTP.  \\
In this case, only the confidential user data is encrypted but not the user credential asset. This condition could lead to the following threat:  An attacker could intercept the credential data during the wireless transmission to the cell tower. 

Moreover, the attacker could use them to impersonate the user and spoof the offered service to obtain the confidential data of the user. Subsequently, the attacker may be able to crack the encryption and gain access to the data.  \\
This example is intended to illustrate how the information of the system can be represented in a diagram and how it can be used to uncover potential threats. \\
The two following sections present the formal representation of the content-specification and the advanced data-flow diagram.

\subsubsection{Formal Content-Specification}
\label{sec:content_specification}
A modeled diagram consists of elements, connectors, assets, and boundaries. Each of these diagram components has an assigned type. The diagram component types and properties are predefined in stencils. The predefined stencils form the so-called content-specification. Its type determines the properties that can be set on an element, connector, or asset. The specification tells which properties are allowed for each type.\\

Furthermore, the types of the diagram components can be assigned in a two-level hierarchy. This means that the diagram components each have an top-type and a sub-type. To illustrate this, consider the previous diagram example. The "Mobile Phone" element contains an element called "Application". "Application" is a sub-type of the top type "Software". The top type "Software" contains several sub-types, for example: "Application", "Operating System" and "Firmware". Sub-types can be used to specify diagram components more precisely. Previously it was pointed out that the properties are assigned to the respective types. Sub-types always contain all properties of their top types, but can also have additional properties. \\
As stated before: the specification can be seen as the bridge between the system-model and the threat-model. A valid diagram can only display content that is defined in the specification, and an anti-pattern can only query what can be modeled. \autoref{sec:relation_Diagram_spec} presents the validation of a diagram using this specification and \autoref{sec:relation_rule_spec} the validation of an anti-pattern.\\

 The specification $S$ is defined as a tuple $S = \langle \Lab, \Z, \B, \T, \K, \V, \Hs, \E, \iota_C, \eta, \gamma \rangle$ where:
 \begin{itemize}
 \setlength\itemsep{0.3cm}
    \item $\Lab$ is a finite set of element types.\\
    All elements in a diagram can only have an assigned type $l \in \Lab$.
    
    \item $\Z$ is a finite set of asset types.\\
    All assets in a diagram can only have an assigned type $z \in \Z$.
    
    \item $\B$ is a finite set of boundary types.\\
    All boundaries in a diagram can only have an assigned type $b \in \B$.
    
    \item $\T$ is a finite set of element connector types.\\
    All connectors in a diagram can only have an assigned type $t \in \T$.
    
    \item $\K$ is a finite set of property keys.\\
    Elements, connectors, and assets in a diagram can have properties. $\K$ contains the predefined property keys.
    
    \item $\V$ is a finite set of property values.\\
    Each property key has a set of possible values. $\V$ contains all possible values that can be assigned to various property keys.
    
    \item $\iota_{\Lab}:$ $\Lab \rightarrow  \mathfrak{P}^\Lab$ assigns each "Top-Level" element component type its "Sub-Level" element component types. \\
    The function $\iota_C$ has an index $C$, which specifies the considered set of diagram component types. In this case only the element component types are considered. Moreover, the function $\iota_{\Lab}$ is \emph{undefined} for each "Sub-Level" element component type since there are only two levels of hierarchy, mathematically expressed: \\
    $\forall l \in \Lab , \big(( \exists l' \in  \Lab, l \in \iota_{\Lab}(l')) \implies \iota_{\Lab}(l) = \emph{undefined} \big) $
    
     \item $\iota_{\Z}:$ $\Z \rightarrow  \mathfrak{P}^\Z$ assigns each "Top-Level" asset component type its "Sub-Level" asset component types. The definition of this function is similar to $\iota_{\Lab}$.
    
     \item $\iota_{\B}:$ $\B \rightarrow  \mathfrak{P}^\B$ assigns each "Top-Level" boundary component type its "Sub-Level" boundary component types. The definition of this function is similar to $\iota_{\Lab}$.
     
    \item $\iota_{\T}:$ $\T \rightarrow  \mathfrak{P}^\T$ assigns each "Top-Level" connector component type its "Sub-Level" connector component types. The definition of this function is similar to $\iota_{\Lab}$.
    
    \item $\eta$: $(\Lab \cup \Z \cup \T) \rightarrow  \mathfrak{P}^\K$ assigns each element, asset or connector type a set of property keys.
    
    \item $\gamma$: $\K \rightarrow \mathfrak{P}^\V$ assigns each property key a set of possible values.

 \end{itemize}

\subsubsection{Example Content-Specification}
\label{sec:example_content_specification}
This section shows a content-specification to which the example diagram from \autoref{sec:diagram_example} corresponds. Normally, the order of the steps is reversed because the content-specification must already be defined before a diagram can be created. The specification for the example from \autoref{fig:mobile_phone_diagram} is presented below:\\
$S = \langle \Lab, \Z, \B, \T, \K, \V, \Hs, \E, \iota_C, \eta, \gamma \rangle$ 

\begin{itemize}
\setlength\itemsep{0.3cm}
    \item $\Lab = \{ \text{External Interactor, Mobile Phone, Software, Application, Cell Tower,}$ \\ 
     \hspace*{1cm} $\text{Server, Interface, REST Interface, Memory, Database} \}$ \\
    Contains the available element types.
    
    \item $\Z = \{ \text{Confidential Data, User Credentials, User Data} \}$ \\
    Contains the available asset types.
    
    \item$\B = \{\text{Generic Boundary, Untrusted Environment, Trusted Environment}\}$ \\
    Contains the available boundary types.
    
    \item $\T = \{ \text{Generic Connector, Wired, Wireless} \}$ \\
    Contains the available connector types.
    
    \item $\K$ = \{OS, Vendor, Input Validation, Input Sanitization, \\
    \hspace*{1cm} Encrypted, Protocol \}\\
    Contains the available property keys for elements as well as for connectors.
    
    \item $\V = \{ \text{ Unknown, Yes, No, HTTP, HTTPS, Third Party, }$ \\
     $\text{ \hspace{1cm} Own Premise, Android, IOS} \}$ \\
    Contains the available values which can be assigned to property keys.
    
    \item $\iota_{\Lab}(\text{External Interactor}) = \{ \text{Mobile Phone, Cell Tower }\}$\\           
        $\iota_\Lab(\text{Software}) = \{  \text{Application} \}$ \\
        $\iota_\Lab(\text{Server}) = \{  \} $\\
        $  \iota_\Lab(\text{Memory}) = \{  \text{Memory} \} $\\
        $  \iota_\Lab(\text{Interface}) = \{  \text{REST Interface} \}$ \\
    The partial function $\iota_{\Lab}$ describes the hierarchical levels of element components types $\Lab$. For the example it is defined that the element type "Software" is the "Top-Type" of the "Sub-Type": "Application". The "Top-Type" "Sensor" has no assigned "Sub-Types", but it is defined as a "Top-Type". Therefore it returns not \emph{undefined} but the $\emptyset$.

    \item $\iota_\Z(\text{Confidential Data}) = \{  \text{User Credentials, User Data}\} $ \\
    The partial function $\iota_{\Z}$ describes the hierarchical levels of asset components types $\Z$. For the example it is defined that the asset type "Confidential Data" is the "Top-Type" of the "Sub-Types": "User Credentials" and "User Data". 
    
    \item $\iota_\B(\text{Generic Boundary}) = \{  \text{Untrusted Environment, Trusted Environment}\} $ \\
    The partial function $\iota_{\B}$ describes the hierarchical levels of boundary components types $\B$.
    
    \item $\iota_\T(\text{Generic Connector}) = \{  \text{Wired, Wireless}\} $ \\
    The partial function $\iota_{\T}$ describes the hierarchical levels of connector components types $\B$.
    
    \item $\eta(\text{Mobile Phone}) = \{  \text{OS} \} ,  \ \eta(\text{Mobile Phone}) = \{  \text{Vendor} \},\\
        \eta(\text{Application}) = \{  \text{Vendor} \}, \\
        \eta(\text{Server}) = \{  \text{Vendor} \}, \\
        \eta(\text{REST Interface}) = \{  \text{Input Validation} \}, \\
        \eta(\text{REST Interface}) = \{  \text{Input Sanitization} \},\\
        \eta(\text{Database}) = \{\text{Encrypted} \}, \\
        \eta(\text{Wired}) = \{\text{Protocol} \},\\
        \eta(\text{Wireless}) = \{\text{Protocol} \},\\
        \eta(\text{User Credentials}) = \{\text{Encrypted} \},\\
        \eta(\text{Confidential User Data})= \{\text{Encrypted} \}$\\
    The function $\eta$ assigns each element, asset or connector type their available property keys. For example each instance of an element with the assigned type "Database" has the property "Encrypted".
    
    \item $\gamma(\text{OS}) = \{  \text{Unknown, Android, IOS} \},  \\
            \gamma(\text{Vendor}) = \{  \text{Unknown, Third party, Own Premise} \}, \\
           \gamma(\text{Protocol}) = \{  \text{Unknown, HTTP, HTTPS} \}, \\
           \gamma(\text{Encrypted}) = \{  \text{Unknown, Yes, No} \}, \\
           \gamma(\text{Input Validation}) = \{  \text{Unknown, Yes, No} \}$ \\
    The function $\gamma$ assigns each property key its available values.\\
    For example the property "Encrypted" can have exactly one value out of the set "\{Unknown, Yes, No\}".
    
\end{itemize}

\clearpage

\subsubsection{Formal Diagram-Definition}
\label{sec:formal_Diagram_representation}
\autoref{sec:model_notation} discussed the diagram components  and \autoref{sec:diagram_example} demonstrated their intended use. This section contains a formal definition of system-models that were previously only represented graphically. Since this definition does not refer to a specific diagram type, the analysis does support not only diagrams as presented in \autoref{sec:diagram_example} but also other types of diagrams that can be represented in terms of the definition below.

 The definition of a diagram $D$ is defined as a tuple \\
$D = \langle \N, \Y, \A, \R,source,target,\lambda_C,\mu,\delta, \kappa, \rho \rangle$  where:

\begin{itemize}
\setlength
    \itemsep{0.3cm}
    
    \item $\N$ represents the elements in a defined diagram $D$. \\
    Each element in a diagram $D$ has a unique identifier from $\N$.
    
    \item $\Y$ represents the assets in a defined diagram $D$. \\
    Each asset inside a diagram $D$ has a unique identifier from $\Y$.
    
    \item $\A$ represents the boundaries in a defined diagram $D$. \\
    Each boundary inside a diagram $D$ has a unique identifier from $\A$.
    
    \item $\R$ represents the connectors in a defined diagram $D$.  \\
    Each connector in a diagram $D$ has a unique identifier from $R$.The set $R$ can also be $\emptyset$, as not all diagrams contain a connector.
    
    \item $source: \R \rightarrow \N$ is a total function that maps each connector to its \textit{source} element.
    Each connector in a diagram must always have a starting point. 
    
    \item $target: \R \rightarrow \N$ is a total function that maps each connector to its \textit{target} element.
    Each connector in a diagram must always have an endpoint.
    
    \item $\lambda_{\N}: \N \rightarrow \Lab$ is a total function that can map each element identifier to an element type. The function $\lambda_C$ has an index $C$ which defines which set of diagram components is assigned a type. The set $\N$ contains all element identifiers which in turn can only be assigned element types from the set $\Lab$.
    
    \item $\lambda_{\Y}: \Y \rightarrow \Z$ is a total function that maps each asset identifier to an asset type. The definition of this function is similar to $\lambda_{\N}$.
    
    \item $\lambda_{\A}: \A \rightarrow \B$ is a total function that maps each boundary identifier to a boundary type. The definition of this function is similar to $\lambda_{\N}$.
    
    \item $\lambda_{\R}: \R \rightarrow \T$ is a total function that maps each connector
    identifier to  connector type. The definition of this function is similar to $\lambda_{\N}$.
    
    \item $\mu: (\N \cup \Y \cup \R) \times \K \nrightarrow \V$ is a partial function that maps an element, asset  or a connector identifier and a property key to a certain value.\\
    As described before, each element, asset, or connector has an assigned type. Each type can contain multiple properties. Each property has multiple potential values. The function $\mu$ defines the selected value for a property key. It is a partial function as not every property is defined for every component.
      
    \item $\delta \subseteq \N \times \N$ is a transitive, irreflexive, asymmetric relation between elements in $D$. This indicates that one element is contained by another element. The tuple $(n_1,n_2)$ indicates that $n_1$ is the \textit{parent} of $n_2$. So to say $n_1$ \emph{contains} $n_2$ and $n_2$ \emph{is contained by} $n_1$.
    
    \item $\kappa \subseteq \A \times (\A \cup \N)$ is a transitive, irreflexive, asymmetric relation between boundaries and elements in $D$. That indicates that one element is located inside a boundary or a boundary is contained by another boundary. The tuple $(b_1,n_1)$ indicates that the boundary with the identifier $b_1$ contains the element with the identifier $n_1$. An element cannot contain a boundary. However, a boundary can contain other boundaries.
    
    \item $\rho \subseteq (\N \cup \R) \ \times \ \Y$ is a relation between elements and connectors on the one hand and assets on the other hand. This relation indicates which asset is held by an element or connector. Multiple elements and connectors can hold the same asset to represent the movement of the asset through the system. 
\end{itemize}

\clearpage

\subsubsection{Example Diagram-Definition}
To provide a better understanding of this definition, the example in \autoref{sec:diagram_example} is transferred to: $D = \langle \N, \Y, \A, \R,source,target,\lambda_C,\mu,\delta, \kappa, \rho \rangle$:

\begin{itemize}
\setlength\itemsep{0.3cm}

    \item $\N = \{ n_1, n_2, n_3, n_4, n_5 , n_6\}$ \\
    Contains all the unique element identifiers.
    
    \item $\Y = \{ y_1, y_2\}$ \\
    Contains all the unique asset identifiers.
    
    \item $\A = \{a_1,a_2\}$ \\
    Contains all the unique boundary identifiers.
    
    \item $\R = \{ r_1, r_2, r_3, r_4, r_5, r_6\}$\\
    Contains all the unique connector identifiers.
    
    \item $source(r_1) = n_2, source(r_2) = n_4, source(r_3) = n_5, source(r_4) = n_6, source(r_5) = n_5, source(r_6) = n_3$\\
    The \textit{source} function maps each connector identifier to an element identifier since each connector needs a starting point. 
    
    \item $source(r_1) = n_3, source(r_2) = n_5, source(r_3) = n_6, source(r_4) = n_5, source(r_5) = n_3, source(r_6) = n_2$\\
    The \textit{target} function maps each connector identifier to an element identifier since each connector needs an end point. 
    
.

    \item   $\lambda_{\N}(n_1) = \text{Mobile Phone}$, 
            $\lambda_{\N}(n_2) = \text{Application}$, \\ 
            $\lambda_{\N}(n_3) = \text{Cell Tower}$,  
            $\lambda_{\N}(n_4) = \text{Server}$,\\ 
            $\lambda_{\N}(n_5) = \text{REST Interface}$,
            $\lambda_{\N}(n_6) = \text{Database}$\\
            
        The function $\lambda_{\N}$ maps each element instance to a type from $\Lab$. For example, the element with the identifier $n_1$ has the assigned type "Mobile Phone".
        
    \item   $\lambda_{\Y}(y_1) = \text{User Credentials}$,
            $\lambda_{\Y}(y_2) = \text{Confidential User Data}$\\ 
        The function $\lambda_{\Y}$ maps each asset instance to a type from $\Z$. For example, the asset with the identifier $y_1$ has the assigned type "User Credentials".
        
    \item  $\lambda_{\A}(a_1) = \text{Untrusted Environment}, \lambda_{\A}(a_2) = \text{Trusted Environment}$  \\ 
        The function $\lambda_{\A}$ maps each boundary instance to a type from $\B$. For example, the boundary with the identifier $a_1$ has the assigned type "Untrusted Environment".

    \item $\lambda_{\R}(r_1) = \text{Wireless}$, 
    $\lambda_{\R}(r_2) = \text{Wired}$,  \\
    $\lambda_{\R}(r_3) = \text{Wired}$
    $\lambda_{\R}(r_4) = \text{Wired}$ \\
    $\lambda_{\R}(r_5) = \text{Wired}$ 
    $\lambda_{\R}(r_6) = \text{Wireless}$ \\
    The function $\lambda_{\R}$ maps each connector instance to a type from $\T$. For example, the connector with the identifier $r_1$ has the assigned type "Wireless".
    
        \item   $\mu(n_1,\text{Operating System}) = \text{Android}$,
            $\mu(n_1,\text{Vendor}) = \text{Third Party}$, \\
            $\mu(n_2,\text{Vendor}) = \text{Third Party}$, 
            $\mu(n_4,\text{Vendor}) = \text{Third Party}$, \\
            $\mu(n_5,\text{Input Validation}) = \text{Yes}$,
            $\mu(n_5,\text{Input Sanitization}) = \text{No}$, \\
            $\mu(n_6,\text{Encrypted}) = \text{Yes}$, \\
            $\mu(r_1,\text{Protocol}) = \text{HTTP}$,
            $\mu(r_2,\text{Protocol}) = \text{HTTP}$,\\
            $\mu(r_3,\text{Protocol}) = \text{HTTP}$,
            $\mu(r_4,\text{Protocol}) = \text{HTTP}$,\\
            $\mu(r_5,\text{Protocol}) = \text{HTTP}$,
            $\mu(r_6,\text{Protocol}) = \text{HTTP}$,\\
            $\mu(y_1,\text{Encrypted}) = \text{No}$,
            $\mu(y_2,\text{Encrypted}) = \text{Yes}$ \\
            
            The function $\mu$ maps for each element, asset, or connector instance a property key to a concrete value. The function $\mu$ is only valid if the assigned element, asset, or connector type allows the property key and the property value is allowed for the specific property key
    
    \item $\delta =  \{(n_1,n_2),(n_4,n_5),(n_4,n_6)\}$ \\
    The relation $\delta$ describes which element contains another element or which element is contained by another one. For example the element with the identifier $n_1$ contains the element $n_2$. Inside the diagram example, this refers to the "Mobile Phone", which contains the "Application" element.
    
    \item $\kappa  = \{(a_1,n_1), (a_1,n_2), (a_1,n_3), (a_2,n_4), (a_2, n_5), (a_2, n_6)\}$ \\
    The relation $\kappa$ describes which boundary contains another boundary or element. For Example the boundary $a_1$ contains the elements $n_1$, $n_2$ and $n_3$.
     
    \item $\rho =  \{(r_1,y_1), (n_3, y_1), (r_2, y_1), (n5, y_1), (r_3, y_1), (n_6, y_1)$ \\
    \hspace*{1cm} $(r_6,y_2), (n_3, y_2), (r_5, y_2), (n5, y_2), (r_4, y_2), (n_6, y_2) \}$ \\
    The relation $\rho$ describes which element or connector holds an asset. An example tuple $(n_6,y_2)$ from \autoref{fig:mobile_phone_diagram} indicates that the element "Database" $n_6$ \emph{holds} the "User Data" asset $y_2$.
    
\end{itemize}

\clearpage

\subsubsection{Relation between Content-Specification and Diagram-Definition}
\label{sec:relation_Diagram_spec}
This section defines the conformance relation between the diagram 
$D = \langle \N,\allowbreak \Y,\allowbreak \A,\allowbreak \R,\allowbreak source,\allowbreak target,\allowbreak \lambda_C,\allowbreak \mu,\allowbreak \delta,\allowbreak \kappa,\allowbreak \rho \rangle$ and the content-specification $S = \langle \Lab,\allowbreak \Z,\allowbreak \B,\allowbreak \T,\allowbreak \K,\allowbreak \V,\allowbreak \Hs,\allowbreak \E,\allowbreak \iota_C,\allowbreak \eta,\allowbreak \gamma \rangle$. \\

The expression $D \models S $ means that a diagram $D$ conforms to the specification $S$. The symbol $\models$ describes a \emph{conformance relation} and says that $D$ \emph{conforms to} $S$. $D \models S $ holds iff all of the following conditions hold:\\

\begin{itemize}
    \setlength\itemsep{0.3cm}
    \item  $|\N| > 0$   \\
    $D$ contains at least one element.
    
    \item  $ |\R| = 0 \vee (|\R| > 0 \rightarrow |\N | \geq 2)$   \\
    $D$ contains no connector, or if it does, it also contains at least two elements.
    
    \item  $ |\Y| = 0 \vee (|\Y| > 0 \rightarrow (|\N| \geq 1 \vee |\Y| \geq 1))$   \\
    $D$ contains no asset or if it does it also contains at least one element or one connector, because an asset must be held by an element or connector.
    
    \item $\forall n \in \N, \lambda_{\N}(n) \in \Lab$ \\
    Each element $n$ in $D$ must have an assigned type from set $\Lab$.
    
    \item $\forall y \in \Y, \lambda_{\Y}(y) \in \Z$ \\
    Each asset $y$ in $D$ must have an assigned type from set $\Z$.
    
    \item $\forall a \in \A, \lambda_{\A}(a) \in \B$ \\
    Each boundary $a$ in $D$ must have an assigned type from set $\B$.
    
    \item $\forall r \in \R,  \lambda_{\R}(r) \in \T$  \\
    Each connector $r$ in $D$ must have an assigned type from set $\T$.
    
    \item $\forall n \in \N, k \in \K, \mu(n,k) \text{ is } \textit{defined} \Leftrightarrow k \in \eta(\lambda(n))$ \\
    If an element instance $n$ has an assigned property key $k$, then $k$ must be valid for the assigned type of $n$.
    
    \item $\forall y \in \Y, k \in \K, \mu(y,k) \text{ is } \textit{defined} \Leftrightarrow k \in \eta(\chi(y))$ \\
    If an asset instance $y$ has an assigned property key $k$, then $k$ must be valid for the assigned type of $n$.
    
    \item $\forall r \in \R, k \in \K, \mu(r,k) \text{ is } \emph{defined} \Leftrightarrow k \in \eta(\tau(r))$ \\
    If a connector instance $r$ has an assigned property key $k$, then $k$ must be valid for the assigned type of $r$.
    
    \item $\forall j \in (\N \cup \Y \cup \R), k \in \K, \mu(j,k) \text{ is } \emph{defined} \Leftrightarrow \mu(j,k)  \in \gamma(k)$ \\
     If an element, asset, or connector $j$ has a property key $k$, then the assigned value for the property $k$ must be in the set of the possible values for $k$.
   
\end{itemize}

 Every diagram $D$, such that $D \models S $, can be checked against the threat-model, formalized in the next section.

\subsubsection{The Rule-based Threat-Model}
\label{sec:rule_definition}
The following sections describe the rule-based threat-model. A rule consists of a title, a description, an assigned threat type, an impact estimation, a likelihood estimation, and, most importantly, the so-called anti-pattern. \\
The anti-pattern is at the heart of every rule. It expresses threats in a human as well as machine-readable language. Each anti-pattern describes an undesirable condition inside a system.

\subsubsection{Rule Syntax}
\label{sec:rule_syntax}
\autoref{fig:pattern_syntax} and \autoref{fig:filter_syntax} display the full syntax of the context-free anti-pattern grammar. The notation of the grammar was inspired by the "Extended Backus–Naur form" (EBNF), where each line represents a production rule of the syntax. Each production rule, in the following referred to as "term", consists of terminal and non-terminal tokens. A terminal is marked in red and describes an immutable part of the syntax.  The non-terminal tokens of the syntax are marked in blue. Each non-terminal must be replaced by the associated term. The syntax differentiates between "patterns" and "filters". A pattern relates to a diagram component such as the element-pattern (\blue{elPat}).\\
A pattern examines whether an element, asset, boundary, or connector exists within the diagram or not. An assigned type can further restrict each pattern except the flow-pattern. The assigned type of a diagram component can be examined using a type-filter (\blue{typeFil}). The type-filter as well as most of the other filters are optional.  \\
A pattern can but does not necessarily have additional filters assigned. A filter specifies additional conditions that a pattern must meet in order to correspond to a threat. For example, the property-filter (\blue{propFil}) can be applied to an element-pattern, connector-pattern (\blue{conPat})  or asset-pattern (\blue{assetPat}) and verifies whether the component has a specific property with a specific value. The meaning and intended use of all patterns and filters are presented in \autoref{sec:relation_rule_spec}. \\
Some lines of the syntax contain the following black symbols "\mymid", " + " and question mark " ? ".
The black vertical line describes a logical OR. The red vertical bar is a terminal symbol of the language itself.
The plus indicates that the symbols enclosed in black brackets must appear at least once. However, they may also occur multiple times. All red brackets are part of the syntax itself.
The question mark indicates that the preceding symbols can occur but do not need to. 

\setcounter{equation}{0}
\begin{figure}[htb]%
  \fbox{%
    \parbox{\insidefboxwidth}{%
      \abovedisplayskip=0pt\belowdisplayskip=0pt%
      \begin{align*}
        &\blue{query}            &::=& \ \blue{query} \ (\red{\&} \ \blue{query})+ \mymid \red{(}  \blue{query} (\red{\mymid}  \blue{query})+ \red{)} \mymid \blue{pattern}\\
        &\blue{pattern}          &::=& \ \blue{elPat} \mymid \blue{boundPat} \mymid \blue{conPat} \mymid \blue{flowPat}  \\
        &\blue{elPat}            &::=& \ \red{(} \blue{elPat} \ (\red{\mymid} \ \blue{elPat})+  \red{)} \\
        &                        & & \ \ \mymid \ \red{Element} \ (\blue{typeFil}_{\Lab})? \ (\red{\{} \blue{elPatFil} \red{\}})? \\
        &\blue{assetPat}         &::=& \ \red{(} \blue{assetPat} \ (\red{\mymid} \ \blue{assetPat})+  \red{)} \\
        &                        & & \ \ \mymid \ \red{Asset} \ (\blue{typeFil}_{\Z})? \ (\red{\{} \blue{assetPatFil} \red{\}})? \\
        &\blue{boundPat}         &::=& \ \red{(} \blue{boundPat} \ (\red{\mymid} \ \blue{boundPat})+  \red{)} \\
        &                        & & \ \ \mymid \ \red{Boundary} \ (\blue{typeFil}_{\B})? \ (\red{\{} \blue{boundPatFil} \red{\}})? \\
        &\blue{conPat}           &::=& \ \red{(} \blue{conPat} \ (\red{\mymid} \ \blue{conPat})+  \red{)} \\
        &                        & & \ \ \mymid \ \red{Connector} \ (\blue{typeFil}_{\T})? \ \red{\{} \blue{srcFil} \ \red{\&} \ \blue{tgtFil} \ (\blue{\red{\&} \ conPatFil})? \  \red{\}}  \\
        &\blue{flowPat}          &::=& \ \red{(} \blue{flowPat} \ (\red{\mymid} \ \blue{flowPat})+  \red{)} \\
        &                        & & \ \ \mymid \ \red{Flow} \ \red{\{} \blue{srcFil} \ \red{\&} \ \blue{tgtFil} \ (\blue{\red{\&} \ flowPatFil})?   \red{\}} \\
        & \blue{typeFil}_{C}         &::=& \  (\red{!=})? \ \red{"} q \red{"} \ \ \ \ q \ \in \ C \\
        &                        &  & \ \ \mymid (\red{in} \mymid \red{not in})  \ \red{[} \red{"}q_1\red{"} \ (\red{,} \ \red{"}q_i\red{"})* \red{]}  \ \ \ \ \forall q_i \ \in \ C \\
        &\blue{srcFil}           &::=& \ \red{Source} \ \blue{elPat}\\
        &\blue{tgtFil}           &::=& \ \red{Target} \ \blue{elPat} 
      \end{align*}%
    }%
  }
  \caption{Syntax of Patterns}
  \label{fig:pattern_syntax}
\end{figure}

\clearpage

\setcounter{equation}{0}
\begin{figure}[htb]%
  \fbox{%
    \parbox{\insidefboxwidth}{%
      \abovedisplayskip=0pt\belowdisplayskip=0pt%
        \begin{align*}
            &\blue{elPatFil}            &::=& \ \blue{elPatFil} \ (\red{\&} \ \blue{elPatFil})+ \mymid \red{(}  \blue{elPatFil} (\red{\mymid}  \blue{elPatFil})+ \red{)} \\
            &                          && \ \mymid \blue{propFil} \mymid \blue{assetFil}  \mymid \blue{elRelFil} \mymid \blue{conFil} \mymid \blue{flowFil}  \\
            &\blue{assetPatFil}         &::=& \ \blue{assetPatFil} \ (\red{\&} \ \blue{assetPatFil})+ \\
            &                           && \ \mymid \red{(}  \blue{assetPatFil} (\red{\mymid}  \blue{assetPatFil})+ \red{)} \mymid \blue{propFil} \\
            &\blue{boundPatFil}         &::=& \ \blue{boundPatFil} \ (\red{\&} \ \blue{boundPatFil})+ \\
            &                           && \ \mymid \red{(}  \blue{boundPatFil} (\red{\mymid}  \blue{boundPatFil})+ \red{)} \mymid \blue{boundRelFil} \\
            &\blue{conPatFil}           &::=& \ \blue{conPatFil} \ (\red{\&} \ \blue{conPatFil})+ \mymid \red{(}  \blue{conPatFil} (\red{\mymid}  \blue{conPatFil})+ \red{)} \\
            &                           && \ \mymid \blue{propFil} \mymid \blue{assetFil} \mymid \blue{conCrossesFil}\\
            &\blue{flowPatFil}          &::=& \ \blue{flowPatFil} \ (\red{\&} \ \blue{flowPatFil})+ \mymid \red{(}  \blue{flowPatFil} (\red{\mymid}  \blue{flowPatFil})+ \red{)} \\
            &                           && \ \mymid \blue{includesFil} \mymid \blue{flowCrossesFil}\\
            &\blue{tvFil}               &::=& \ \red{"} k \red{"} \ (\red{=} \mymid \red{!=}) \red{"} v \red{"} \ \ \ \ k \in \K, v \in \V  \\
            &                           && \ \mymid \red{"} k\red{"}  \ (\red{in} \mymid \red{not in}) \ \red{[} \red{"}v_1\red{"} \ (\red{,} \ \red{"}v_i\red{"})* \red{]}  \ \ \ \ k \in \K, \forall v_i \in \V \\
            &\blue{assetFil}            &::=& \ \red{Holds} \ \blue{assetPat}\\
            &\blue{elRelFil}            &::=& \ \red{Contains} \  (\red{no})? \ \blue{elPat} \\
            &                           & & \mymid (\red{Not})? \ \red{Contained by} \ (\blue{elPat} \mymid \blue{boundPat})  \\
            &\blue{boundRelFil}            &::=& \ \red{Contains} \  (\red{no})? \ (\blue{elPat} \mymid \blue{boundPat}) \\
            &                           & & \mymid (\red{Not})? \ \red{Contained by} \ \blue{boundPat} \\
            &\blue{conFil}            &::=& \ \red{Has} \ (\red{no})? \ \red{Connector} \ (\red{"} \blue{conType}\red{"})? \\
            &                           & &\hspace{0.5cm} \red{\{} (\blue{srcFil} \mymid\blue{tgtFil}) \ (\red{\&} \ \blue{conPatFil})?   \red{\}} \\
            &\blue{flowFil}           &::=& \red{Has} \ (\red{no})? \ \red{Flow} \ \red{\{} (\blue{srcFil} \mymid\blue{tgtFil}) \ (\red{\&} \ \blue{flowPatFil})?   \red{\}} \\
            &\blue{conCrossesFil}       &::=& \ \red{Crosses} \ (\blue{elPat} \mymid \blue{boundPat})\\
            &\blue{flowCrossesFil}      &::=& \ \red{Crosses} \ (\blue{elPat} \mymid \blue{boundPat})\\
            &\blue{includesFil}         &::=& \ \red{Includes} \ (\red{no} \mymid \red{only})? \ (\blue{elPat} \mymid \blue{conPat}) \ \
        \end{align*}%
    }%
  }
  \caption{Syntax of Filters}
  \label{fig:filter_syntax}
\end{figure}

\clearpage

\subsubsection{Flow Definition}
\label{sec:flow_definition}
Before the syntax is explained in the next section, the concept of \emph{flows} is introduced in this section. A flow in a diagram is a finite alternating sequence of elements and connectors between a start point (element) and an endpoint (element).  It should also be mentioned that these two points do not have to be directly connected to each other. In addition, a flow cannot contain loops so that an element can appear several times in the flow sequence. In order to prevent infinite loops, each connector and connector must be unique within a flow sequence.\\ 
A flow $\ p = seq(\N \ \cup \ \R)$ has the following structure $p = (n_1,r_1,n_2, \ldots, r_{i-1},n_i)$. The following functions can be applied to a flow:\\

\begin{itemize}
    \setlength\itemsep{0.3cm}
    
    \item $\mathsf{pSource}(p)$ is a function that takes a flow sequence $p$ as an argument and returns the first element identifier in the sequence: \\ $\mathsf{pSource}(n_1,r_1 \ldots r_{i-1},n_i) =  n_1$. 
    
    \item $\mathsf{pTarget}(p)$ is a function that takes a flow sequence $p$ as argument and returns the last element identifier in the sequence:\\
    $\mathsf{pTarget}(n_1,r_1 \ldots r_{i-1},n_i) =  n_i$. 
    
    \item $\mathsf{elements}(p)$ is a function that takes a flow sequence $p$ as argument and returns the set of all element identifiers in  the flow:\\ $\mathsf{elements}(n_1,r_1 \ldots r_{i-1},n_i)  =  \{n_1 \ldots n_i\}$. 
    
    \item $\mathsf{connectors}(p)$ is a function that takes a flow $p$ as argument and returns the set of all connector identifiers in  the flow:\\
    $\mathsf{connectors}(n_1,r_1 \ldots r_{i-1},n_i)  =  \{r_1 \ldots r_{i-1}\}$. 
\end{itemize}

The function $\mathsf{flows}(D,src,tgt)$ takes three arguments: a diagram instance $D$, a source element $src$ and target element $tgt$ identifier. The function returns a set of all possible flow sequences between the source element, and the target element within the diagram. Each of these sequences represents a single flow $p$ in $D$. At this point it should be emphasized that mathematically a flow can be compared with a path between two elements. The formal definition of $\mathsf{flows}(D,src,tgt) = $:\\

\setcounter{equation}{0}
\begin{align} 
    \label{eq:altseq}\{(n_1,r_1 \ldots r_{i-1},n_i) \ | \ &  (\forall 1 {\leq} j {<} i,\ source(r_j) = n_{j}  \wedge \ target(r_j) = n_{j+1}   \\
     \label{eq:noloop}&  \wedge (\nexists\ 1 {\leq} j{<}k < i, \ (n_j = n_k ) \}
\end{align}
Line (\ref{eq:altseq}) states that all connectors within a flow always link two consecutive elements and the target of connector $r_j$ is always the source of the connector $r_{j+1}$.\\
The second line (\ref{eq:noloop}) expresses that each element must be unique in the sequence. \\
The corresponding algorithm of this function is presented in \autoref{sec:flows_algorithm}.

\subsubsection{Relation between Analysis Language and Content-Specification}
\label{sec:relation_rule_spec}
This section depicts the conformance relation between the rule anti-pattern content and the content-specification $S = \langle \Lab, \Z, \B, \T, \K, \V, \Hs, \E, \iota_C, \eta, \gamma \rangle$. Therefore, the conformance relation is denoted as $term \models S$, where $term$ refers to a rule in the syntax.\\

In some cases the conformance relation requires a context, denoted as $\models_c$, where $c$ is the context. The context $c \subseteq (\Lab \cup \Z \cup \T \cup \B)$ is a subset of element types, asset types, boundary types or connector types which are under consideration. Since a path pattern (\blue{pathPat}) cannot have an assigned type, it cannot be assigned to the context $c$.   \\

By $term \models S \Leftrightarrow term' \models_{c} S$ it is denoted that $term$ conforms to a specification $S$ iff $term'$ conforms to $S$ under context $c$. Not all terms require a context for their conformance relation, in particular the conformance relation for \blue{query} does not require a context. \nv \ is defined as a special replacement for optional parameters which are omitted. Since \nv \ is not part of the specification $S$, it never conforms to it: $\nv\nvDash S$ \\

Since most of the pattern and filter conformance relations differ only marginally, only the first is described in detail in order to avoid repetition.\\

 The \blue{query} term is the root or starting point of the syntax. This is a non-terminal symbol which can be logically linked ( \red{\&} \mymid \red{\mymid}) multiple times with itself or can be replaced by a \blue{pattern} term. It is the only term that can be used to connect different patterns. 
\begin{align}
\setcounter{equation}{0}
    & \bp \ \blue{query}_1 \ \red{\&} \ \blue{query}_2 \  \models S  \Leftrightarrow \blue{query}_1 \models S \wedge \blue{query}_2 \models S \\
    & \bp \ \red{(} \blue{query}_1 \ \red{\mymid} \ \blue{query}_2 \red{)} \  \models S  \Leftrightarrow \blue{query}_1 \models S \wedge \blue{query}_2 \models S 
\end{align}
Line (1) describes how multiple query terms are linked with a logical AND (\red{\&}). Both sub-terms have to conform to the content-specification to form a valid rule query.   \\
The second line (2) shows two logically OR (\red{\mymid}) connected query terms.  Although the two query terms are linked with a logical OR on the left, the conformance relationships to the right are linked with an AND ($\wedge$). The entire term conforms to the specification $S$ only if the first $\blue{query}_1$ term AND ($\wedge$) the second $\blue{query}_2$ term conform to $S$.\\

The \textbf{element-pattern} \blue{elPat} is used to examine the elements of a diagram. 
\begin{align}
\setcounter{equation}{0}
    & \bp \ \red{(}\blue{elPat}_1  \red{\mymid}  \blue{elPat}_2 \red{)} \ \models \ S \ \Leftrightarrow \ \blue{elPat}_1 \ \models \ S \wedge \blue{elPat}_2 \ \models \ S \\
    & \bp \ \red{Element} \ (\blue{typeFil})? \ (\red{\{} \blue{elPatFil} \red{\}})? \ \models \ S \ \Leftrightarrow \\
    & \hspace{0.5cm} (\blue{typeFil} \models_{\Lab} S \ \vee \blue{typeFil} = \nv ) \  \wedge \ (\blue{elPatFil} = \nv \vee  \blue{elPatFil} \models_{c} S )
\end{align}
The first line (1) of every pattern term describes that the same pattern can occur several times if it is linked by a logical OR (\red{\mymid}). Each sub-pattern has to conform to the specification $S$. Therefore they are connected by a logical AND ($\wedge$).  \\
Line (2) presents the complete element-pattern. Each pattern starts with its corresponding keyword (\red{Element}, \red{Asset}, \red{Boundary}, \red{Connector}, \red{Flow}). It should be noted that the type-filter (\blue{typefil}) and the element-pattern-filters (\blue{elPatFil}) definition are marked with a question mark. Therefore, they can be omitted. If they are omitted, these terms are replaced by \nv. \\
(3) The provided type-filter (\blue{typefil}) is either omitted and thus \nv \ or must conform to the specification $S$. The context of the type-filter evaluation is $\Lab$ since an element-pattern is used to analyze all element-components within the diagram $D$. If an element-pattern has additional filters (\blue{elPatFil}), these have to conform to the specification $S$ under the consideration of the context $c$. This context contains either the set of element-types ($c \subseteq \Lab$) defined within the provided type-filter (\blue{typefil}) or \nv \ if the type-filter was omitted. \\
The conformance relation of the asset-pattern (\blue{assetPat}) and boundary-pattern (\blue{boundPat}) is similar to this one.\\

 The \textbf{asset-pattern} (\blue{assetPat}) is used to examine the assets of a system. It is the only pattern that cannot stand alone but is used in the so-called asset-filter (\blue{assetFil}). 
\begin{align*}
\setcounter{equation}{0}
    & \bp \ \red{(}\blue{assetPat}_1  \red{\mymid}  \blue{assetPat}_2 \red{)} \ \models \ S \ \Leftrightarrow \ \blue{assetPat}_1 \ \models \ S \wedge \blue{assetPat}_2 \ \models \ S \\
    & \bp \ \red{Asset} \ (\blue{typeFil})? \ (\red{\{} \blue{assetPatFil} \red{\}})? \ \models \ S \ \Leftrightarrow \\
    & \hspace{0.5cm} (\blue{typeFil} \models_{\Z} S \ \vee \blue{typeFil} = \nv ) \  \wedge \ (\blue{assetPatFil} = \nv \vee  \blue{assetPatFil} \models_{c} S )
\end{align*}

The \textbf{boundary-pattern} (\blue{boundPat}) examines the defined boundary components of a diagram. 
\begin{align*}
\setcounter{equation}{0}
    & \bp \ \red{(}\blue{boundPat}_1  \red{\mymid}  \blue{boundPat}_2 \red{)} \ \models \ S \ \Leftrightarrow \  \blue{boundPat}_1 \ \models \ S \wedge \blue{boundPat}_2 \ \models \ S \\
    & \bp \ \red{Boundary} \ (\blue{typeFil})? \ (\red{\{} \blue{boundPatFil} \red{\}})? \ \models \ S \ \Leftrightarrow \\
    & \hspace{0.5cm} (\blue{typeFil} \models_{\B} S \ \vee \blue{typeFil} = \nv ) \  \wedge \ (\blue{boundPatFil} = \nv \vee  \blue{boundPatFil} \models_{c} S )
\end{align*}

 The \textbf{connector-pattern} (\blue{conPat}) is used to investigate the connectors.
\begin{flalign}
\setcounter{equation}{0}
    & \bp \ \red{(}\blue{conPat}_1  \red{\mymid}  \blue{conPat}_2 \red{)} \ \models \ S \ \Leftrightarrow \ \blue{conPat}_1 \ \models \ S \wedge \blue{conPat}_2 \ \models \ S \\
    &\bp \ \red{Connector} \ (\blue{typeFil})? \ \red{\{}\blue{srcFil} \ \red{\&} \ \blue{tgtFil} \ (\red{\&} \ \blue{conPatFil})? \red{\}}  \models \ S \ \Leftrightarrow \\
    & \hspace{0.5cm} (\blue{typeFil} \models_{\T} S \ \vee \blue{typeFil} = \nv ) \ \wedge \ \blue{srcFil} \ \models \  S  \   \wedge \ \blue{tgtFil} \ \models \ S \  \wedge\\
    & \hspace{0.7cm} (\blue{conPatFil} = \nv \ \vee \ \blue{conPatFil} \ \models_{c} \ S)
\end{flalign}
Line (2) depicts the complete connector-pattern. Compared to the previous patterns this pattern includes a mandatory source-filter (\blue{srcFil}) and target-filter (\blue{tgtFil}). \\
(3) In order that the connector-pattern conforms to the specification $S$, both the source-filter and the target-filter have to conform to $S$.\\

The \textbf{flow-pattern} (\blue{flowPat}) is used to evaluate these flow sequences. \autoref{sec:flow_definition} introduced the concept of flows.

\begin{flalign}
\setcounter{equation}{0}
    & \bp \ \red{(}\blue{flowPat}_1  \red{\mymid}  \blue{flowPat}_2 \red{)} \ \models \ S \ \Leftrightarrow \ \blue{flowPat}_1 \ \models \ S \wedge \blue{flowPat}_2 \ \models \ S \\
    &\bp \ \red{Flow} \ \red{\{}\blue{srcFil} \ \red{\&} \ \blue{tgtFil} \ (\red{\&} \ \blue{flowPatFil})? \red{\}}  \models \ S \ \Leftrightarrow \\
    & \hspace{0.5cm} \blue{srcFil} \ \models \  S  \   \wedge \ \blue{tgtFil} \ \models \ S \  \wedge \  (\blue{flowPatFil} = \nv \ \vee \ \blue{flowPatFil} \ \models \ S)
\end{flalign}
(2) The flow-pattern is the only pattern that contains no type specification as a flow consists of various elements and connectors. Similar to the connector-pattern (\blue{conPat}), the flow-pattern always includes a source-filter (\blue{srcFil}) and a target-filter (\blue{tgtFil}). \\
(3) The source-filter as well as the target-filter have to conform to $S$. The validation of the additional filters applied to the flow-pattern requires no additional context $c$, as a flow cannot have an assigned type.\\

The \textbf{type-filter} (\blue{typeFil}) is used to examine elements, assets, boundaries and connectors according to their assigned type. Therefore, the evaluation of the type-filter also has a context $c$ which contains the set of diagram component types, depending on the pattern that holds the type-filter.
\begin{flalign}
\setcounter{equation}{0}
    & \bp \ \red{:} \ (\red{!=})?  \ \red{"} q \red{"} \models_c \ S \ \Leftrightarrow q \ \in c \\
    & \bp \ \red{:} \ (\red{in} \mymid \red{not in})?  \ \red{[} \ \red{"}q_1\red{"} \ \red{,} \  \ldots \red{"} q_i \red{"} \red{]}  \ \models_c \ S \ \Leftrightarrow \forall q_i \in c
\end{flalign}
The context $c$ is either the set of element-types ($\Lab$), asset-types ($\Z$), boundary-types ($\B$), or connector-types ($\T$). Depending on the previous pattern. \\
Line (1) describes the simple version of the type-filter. In this case it is checked whether the diagram component under investigation has or has exactly not the assigned  type defined by $q$. The type-filter conforms to the specification $S$ if the specified type is contained by the type set inside the context $c$. \\
Line (2) depicts an advanced version of the type-filter.In this case it is checked whether the diagram component has one of or has exactly none of the specified types defined by $q_1 \ \ldots \ q_i$. The type-filter conforms to the specification $S$ if all the specified types  $q_1 \ \ldots \ q_i$ are contained by the type set inside the context $c$.  \\

The \textbf{source-filter} (\blue{srcFil}) is used within the connector-pattern and flow-pattern to specify the source element. The \textbf{target-filter} (\blue{tgtFil}) is used to specify the target element.

\begin{flalign}
\setcounter{equation}{0}
    & \bp \ \red{Source} \ \blue{elPat}  \ \models_c \ S  \Leftrightarrow \blue{elPat} \models S \\
    & \bp \ \red{Target} \ \blue{elPat}  \ \models_c \ S  \Leftrightarrow \blue{elPat} \models S
\end{flalign}
(1-2) The source-filter as well as the target-filter conforms to the specification if the defined element-pattern (\blue{elPat}) conforms to $S$.\\

 The term \blue{elPatFil} describes all filters which can be assigned to an element-pattern. An element-pattern can be filtered according to its properties (\blue{propFil}), its assets (\blue{assetFil}), its relationships to other elements and boundaries (\blue{elRelFil}) and its incoming and outgoing connectors (\blue{conFil}) as well as flows (\blue{flowFil}). 
\begin{flalign}
\setcounter{equation}{0}
    & \bp \ \blue{elPatFil}_1 \  \red{\&} \  \blue{elPatFil}_2 \ \models_c \ S \ \Leftrightarrow \ \blue{elPatFil}_1 \ \models_c \ S \wedge \blue{elPatFil}_2 \ \models_c \ S \\
    & \bp \ \red{(}\blue{elPatFil}_1  \red{\mymid}  \blue{elPatFil}_2 \red{)} \ \models_c \ S \ \Leftrightarrow \  \blue{elPatFil}_1 \ \models_c \ S \wedge \blue{elPatFil}_2 \ \models_c \ S 
\end{flalign}
Each pattern of the syntax can be assigned multiple filters. These filters are connected either using a logical AND (\red{\&}) or a logical OR (\red{\mymid}). The conformance relation of the following filters is similar to this one. \\
Line (1) depicts two filters that are logically connected by an AND  (\red{\&}). In this case, each of the filters has to conform to the specification $S$. \\
Line (2) depicts two filters that are logically connected by an OR  (\red{\mymid}). Similar to the line (1), each of the filters has to conform to the specification $S$. Since it is not a matter of the semantic evaluation of the language, both conformance relations are connected in line 4 with an AND ($\wedge$). \\

The asset-specific filters are defined using the term \blue{assetPatFil}. An asset-pattern can only be filtered according to its properties \blue{propFil}.
\begin{align*}
\setcounter{equation}{0}
   & \bp \ \blue{assetPatFil}_1 \  \red{\&} \  \blue{assetPatFil}_2 \ \models_c \ S \ \Leftrightarrow \ \blue{assetPatFil}_1 \ \models_c \ S \wedge \blue{assetPatFil}_2 \ \models_c \ S \\
   & \bp \ \red{(}\blue{assetPatFil}_1  \red{\mymid}  \blue{assetPatFil}_2 \red{)} \ \models_c \ S \ \Leftrightarrow \  \blue{assetPatFil}_1 \ \models_c \ S \wedge \blue{assetPatFil}_2 \ \models_c \ S 
\end{align*}

The filters of the boundary-pattern are specified using the \blue{boundPatFil} term. Similar to the asset-filters, a boundary-pattern has only one filter which refers to the relationship between the boundary and other elements \blue{boundRelFil}.
\begin{align*}
\setcounter{equation}{0}
    & \bp \ \blue{boundPatFil}_1 \  \red{\&} \  \blue{boundPatFil}_2 \ \models_c \ S \ \Leftrightarrow \  \blue{boundPatFil}_1 \ \models_c \ S \wedge \blue{boundPatFil}_2 \ \models_c \ S \\
    & \bp \ \red{(}\blue{boundPatFil}_1  \red{\mymid}  \blue{boundPatFil}_2 \red{)} \ \models_c \ S \ \Leftrightarrow \  \blue{boundPatFil}_1 \ \models_c \ S \wedge \blue{boundPatFil}_2 \ \models_c \ S 
\end{align*}

 The connector-specific filters are defined using the term \blue{conPatFil}. A connector-pattern can be filtered according to its assigned properties (\blue{propFil}), its assets (\blue{assetFil}) and whether it crosses an element or boundary (\blue{conCrossesFil}). 
\begin{align*}
\setcounter{equation}{0}
    & \bp \ \blue{conPatFil}_1 \  \red{\&} \  \blue{conPatFil}_2 \ \models_c \ S \ \Leftrightarrow \  \blue{conPatFil}_1 \ \models_c \ S \wedge \blue{conPatFil}_2 \ \models_c \ S \\
    & \bp \ \red{(}\blue{conPatFil}_1  \red{\mymid}  \blue{conPatFil}_2 \red{)} \ \models_c \ S \ \Leftrightarrow \ \blue{conPatFil}_1 \ \models_c \ S \wedge \blue{conPatFil}_2 \ \models_c \ S 
\end{align*}

The flow-related filters are given by the term \blue{flowPatFil}. A flow-pattern can be filtered according to its included elements and connectors (\blue{includesFil}) and whether it crosses a boundary or element (\blue{flowCrossesFil}).
\begin{align*}
\setcounter{equation}{0}
    & \bp \ \blue{flowPatFil}_1 \  \red{\&} \  \blue{flowPatFil}_2 \ \models \ S \ \Leftrightarrow \  \blue{flowPatFil}_1 \ \models \ S \wedge \blue{flowPatFil}_2 \ \models \ S \\
    & \bp \ \red{(}\blue{flowPatFil}_1  \red{\mymid}  \blue{flowPatFil}_2 \red{)} \ \models \ S \ \Leftrightarrow \ \blue{flowPatFil}_1 \ \models \ S \wedge \blue{flowPatFil}_2 \ \models \ S 
\end{align*}

The \textbf{property-filter} \blue{propFil} is used to examine elements, assets and connectors according to their assigned properties.  

\begin{flalign}
\setcounter{equation}{0}
    & \bp  \ \red{"} k \red{"} \ (\red{=} \mymid \red{!=})  \ \red{"} v \red{"} \  \models_c \ S \ \Leftrightarrow \\
    & \hspace{0.5cm} k \in \K \wedge (c = \nv \vee \ \forall 1{\leq}j{\leq}i, k \in \eta(c_j)) \wedge  v \in \gamma(k)  \\
    & \bp \ \red{"} k \red{"}\ (\red{in} \mymid \red{not in})  \ \red{[} \ \red{"}v_1\red{"} \ \red{,} \  \ldots \red{"} v_i \red{"} \red{]}  \ \models_c \ S \ \Leftrightarrow \\
    & \hspace{0.5cm} k \in \K \wedge (c = \nv \vee \ \forall 1{\leq}j{\leq}i, k \in \eta(c_j)) \wedge \forall 1{\leq}j{\leq}i,  v_j \in \gamma(k)  
\end{flalign}
The first two lines (1-2) describe the first variant of this filter. This property-filter conforms to $S$ if $k$ is part of the property key set $ \K $. Moreover, $v$ has to be a valid value for the key $k$ determined by the function $\gamma(k)$. If the context $ c $ is not \nv , \ then the property key $k$ must be valid for each type inside the type set stored in context $c$ determined by the function $\eta(c)$.  \\
The second variant of this filter is described in lines (3-4). Similar to the first variant, the property key $ k $ must be part of $ \K $, and if the context $c$ is not \nv, then it must be a valid key for each component type inside the context. Unlike the first variant, this one deals with a set of values $\{v_1, v_2 \ldots v_i\}$, which must all be valid values for $ k $. \\

The \textbf{asset-filter} (\blue{assetFil}) determines whether an element \red{holds} an asset or a connector transports an asset. The filter can be applied to an element-pattern (\blue{elPat}) or a connector-pattern (\blue{conPat}), since these are the only diagram components which can be related to an asset.
\begin{flalign}
\setcounter{equation}{0}
    & \bp \ \red{Holds} \ \blue{assetPat}  \ \models_c \ S  \Leftrightarrow \blue{assetPat} \models S
\end{flalign}
(1) The asset-filter conforms to the specification if the defined asset-pattern (\blue{assetPat}) conforms to $S$.\\

The \textbf{element-relation-filter} (\blue{elRelFil}) can only be used in an element-pattern (\blue{elPat}) and examines the relationships between an element and other elements and boundaries. The context $c$ is either \nv\ or an element type (\blue{elType}).
\begin{flalign}
\setcounter{equation}{0}
    & \bp \ \red{Contains} \  (\red{no})? \ \blue{elPat}  \ \models_c \ S  \Leftrightarrow \blue{elPat} \models S \\
    & \bp \ (\red{Not})? \ \red{Contained by}  \ \blue{elPat} \ \models_c \ S  \Leftrightarrow \blue{elPat} \models S \\
    & \bp \ (\red{Not})? \ \red{Contained by}  \ \blue{boundPat} \ \models_c \ S  \Leftrightarrow \blue{boundPat} \models S 
\end{flalign}
(1) The first variant of this filter is valid if the defined element-pattern (\blue{elPat}) conforms $S$. \\
The conformance relation of line (2) is similar to the first variant in line (1). \\
Line (3) conforms to the specification if the boundary-pattern (\blue{boundPat}) does.\\
 
The \textbf{boundary-relation-filter} (\blue{boundRelFil}) is the only filter that can be assigned to a boundary-pattern (\blue{boundPat}). This filter is similar to the element-relation-filter, but in this case from the perspective of a boundary. The context $c$ is either \nv \ or a set of boundary types (\blue{boundType}).
\begin{align*}
\setcounter{equation}{0}
    & \bp \ \red{Contains} \ (\red{no})? \ \blue{elPat}  \ \models_c \ S  \Leftrightarrow \blue{elPat} \models S \\
    & \bp \ \red{Contains} \ (\red{no})? \ \blue{boundPat}  \ \models_c \ S  \Leftrightarrow \blue{boundPat} \models S \\
    & \bp \ (\red{Not})? \ \red{Contained by}  \ \blue{boundPat} \ \models_c \ S  \Leftrightarrow \blue{boundPat} \models S 
\end{align*}
The conformance relation of this filter is similar to the element-relation-filter (\blue{elRelfil}). \\

 The \textbf{connector-filter} (\blue{conFil}) can only be assigned to an element-pattern (\blue{elPat}). This filter is very similar to the connector-pattern, but it has either a source-filter (\blue{srcFil}) or a target-filter (\blue{tgtFil}). The element-pattern which contains this filter forms the counterpart of the defined source or target-filter.
\begin{flalign}
\setcounter{equation}{0}
    & \bp \ \red{Has} \ (\red{No})? \ \red{Connector} \ (\blue{typeFil})? \ \red{\{} ( \blue{srcFil} \mymid \blue{tgtFil} ) \ (\red{\&} \ \blue{conPatFil})?   \red{\}} \models_c S  \Leftrightarrow \\
    & \hspace{0.7cm} (\blue{typeFil} \ \models_{\T} \ S  \ \vee \ \blue{typeFil} \ = \ \nv ) \  \wedge \\
     & \hspace{0.8cm}  (\blue{srcFil} \  \models \  S \ \vee \ \blue{tgtFil} \ \models \ S) \ \wedge \\
    & \hspace{0.9cm} (\blue{conPatFil} = \nv  \vee  \blue{conPatFil} \ \models_{c} \ S) 
\end{flalign}
Line (1) depicts the full syntax of the filter. The syntax is similar to the connector-pattern (\blue{conPat}), except that this filter requires either a source-filter or a target-filter. \\
(3) The specified source or target-filter must conform to the specification $S$. The other is \nv \ and does therefore not conform to the specification. \\

The \textbf{flow-filter} (\blue{flowFil}) is similar to the flow-pattern (\blue{flowPat}) but it only requires either the source (\blue{srcFil}) or the target-filter (\blue{tgtFil}). This filter can only be assigned to an element-pattern (\blue{elPat}). The conformance relation of this filter is similar to the connector-filter (\blue{conFil}).
\begin{align*}
\setcounter{equation}{0}
    & \bp \ \red{Has} \ (\red{No})? \ \red{Flow} \ \red{\{} ( \blue{srcFil} \mymid \blue{tgtFil} ) \ (\blue{\red{\&}  flowPatFil})?   \red{\}} \models_c S  \Leftrightarrow \\
    & \hspace{0.5cm} (\blue{srcFil} \  \models \  S \ \vee \ \blue{tgtFil} \ \models \ S) \ \wedge \  (\blue{flowPatFil} = \nv  \vee  \blue{flowPatFil} \ \models \ S) 
\end{align*}

The \textbf{connector-crosses-filter} (\blue{conCrossesFil}) can only be assigned to a connector-pattern (\blue{conPat}) or a connector-filter (\blue{conFil}). It is used to check if a connector crosses a boundary or element.

\begin{flalign}
\setcounter{equation}{0}
    & \bp \ \red{Crosses} \ \blue{elPat} \ \models_c \ S \ \Leftrightarrow \ \blue{elPat} \ \models \ S\\
    & \bp \ \red{Crosses} \ \blue{boundPat} \ \models_c \ S \ \Leftrightarrow \ \blue{boundPat} \ \models \ S
\end{flalign}
(1) The first variant of this filter is valid if the defined element-pattern (\blue{elPat}) conforms $S$. \\
Line (2) conforms to the specification if the boundary-pattern (\blue{boundPat}) does. \\

The \textbf{flow-crosses-filter} (\blue{flowCrossesFil}) can only be assigned to a flow-pattern (\blue{flowPat}) or a connector-filter (\blue{flowFil}). The conformance relation of this filter is similar to the connector-crosses-filter (\blue{conCrossesFil}). It is used to check if one of the connectors within the flow crosses a boundary or element.
\begin{align*}
\setcounter{equation}{0}
    & \bp \ \red{Crosses} \ \blue{elPat} \ \models_c \ S \ \Leftrightarrow \ \blue{elPat} \ \models \ S\\
    & \bp \ \red{Crosses} \ \blue{boundPat} \ \models_c \ S \ \Leftrightarrow \ \blue{boundPat} \ \models \ S
\end{align*}

The \textbf{includes-filter} (\blue{includesFil}) can only be applied to a flow-pattern (\blue{flowPat}) or a flow-filter (\blue{flowFil}). It is used to check if a flow contains a specific element or connector.
\begin{flalign}
\setcounter{equation}{0}
    & \bp \ \red{Includes} \ (\red{no} \mymid \red{only})? \ \blue{elPat} \ \models_c \ S \ \Leftrightarrow \ \blue{elPat} \ \models \ S\\
    & \bp \ \red{Includes} \ (\red{no} \mymid \red{only})? \ \blue{conPat} \ \models_c \ S \ \Leftrightarrow \ \blue{conPat} \ \models \ S
\end{flalign}
(1) The first variant of this filter is valid if the defined element-pattern (\blue{elPat}) conforms to $S$. \\
(2) The second variant of this filter is valid if the defined connector-pattern (\blue{conPat}) conforms to $S$. \\

\clearpage

\subsection{Method Concept}
The previous sections presented the formal system-model and threat-model. The following sections discuss the semantic evaluation of the language and its grammar. The first part elaborates the evaluation of the individual patterns and applied filters. Section  \ref{sec:flows_algorithm} presents the $\mathsf{flows}(D,src,tgt)$ algorithm.

\subsubsection{Semantic Evaluation of Anti-Patterns}
\label{sec:evaluation_patterns_filters}

In contrast to the previous section, the semantic evaluation on the basis of a concrete diagram is examined, and not whether an anti-pattern is syntactically correct.  $D = \langle \N,\allowbreak \Y,\allowbreak \A,\allowbreak \R,\allowbreak source,\allowbreak target,\allowbreak \lambda_C,\allowbreak \mu,\allowbreak \delta,\allowbreak  \kappa,\allowbreak \rho \rangle$ . Therefore, the context $c$ has to be redefined.\\
In this case, the context $c \in (\N \cup \Y \cup \A \cup \R \cup \{seq(\N \cup \R)\}) $  contains an element, asset, boundary or connector identifier or a flow sequence which is under consideration of the evaluation. \\
Only for the type-filter (\blue{typeFil}) has a special context, denoted as capital $C$. This special context is necessary because the type filter considers all diagram components according to the assigned type. In this case the context $C$ contains either the set of all element identifiers ($\N$), asset identifiers ($\Y$),  boundary identifiers ($\A$) or connector identifiers ($\R$). \\

Let  $\sem{term}_{D}$, $\sem{term}_{D,c} or \sem{term}_{C}$ be the evaluation function of a term with respect to a diagram $D$ and an optional context $c$, type-filter context $C$. The evaluation function follows a recursive manner, where each evaluation results in a set of tuples. Each tuple has the form $(x,M)$. Where $x$ represents the element, asset, boundary, connector identifier, or flow sequence, which is under investigation. $M$ is a set containing the additionally affected elements, assets, boundaries, connector identifiers, and flow sequences that result from the applied filters. For the top-level term \blue{query} it is guaranteed that $M$ also contains $x$ itself. This is not true for the evaluation of some of the sub-terms. $x$ is relevant only as an intermediate result for the recursive calls to $\sem{}$. It is not a relevant result of the analysis, which only asks for the affected elements $M$. The symbol "$\_$" is used to indicate that any value could be put at this position of the tuple.\\

The \blue{query} term is the entry point of each rule query evaluation. Each \blue{query} term is replaced by one of the \blue{pattern} terms.
\begin{align}
\setcounter{equation}{0}
    & \bp \ \sem{\blue{query}_1 \ \red{\&} \ \blue{query}_2} = \\
    & \hspace{0.5cm}\{ (\_, M_1 \cup M_2) \mymid (x_1, M_1) \in \sem{\blue{query}_1}_{D} \ \wedge (x_2, M_2) \in \sem{\blue{query}_2}_{D} \} \\
    & \bp \ \sem{\red{(}\blue{query}_1 \ \red{\mymid} \ \blue{query}_2\red{)}}_{D} = \sem{\blue{query}_1}_{D} \ \cup \ \sem{\blue{query}_2}_{D}
\end{align}
(1) The evaluation of several \blue{query} terms, which are connected by a logical AND (\red{\&}), requires that each of these patterns does not return an empty result. If one of the terms returns an empty set, the end result is the empty set. If all terms return a non-empty set, the end result is the cross-product of the results, meaning all combinations of affected elements are returned. Since \blue{query} is the top-level entry point of the evaluation function, the first member of the returned tuple ($\_$) is irrelevant. \\
(2) If multiple \blue{query} terms are linked by a logical OR (\red{\mymid}), the end result of the evaluation is the union of the resulting sets of tuples. A logical OR requires that only one of the two evaluations return a non-empty set. If both return an empty set, then the end result is the empty set. If one evaluation returns a non-empty set and the other does not, then the end result is the non-empty set. If both evaluations return a non-empty set, then the end result is the union of both sets. In this case, the cross product is not needed as the result of each \blue{query} can independently realize the threat.  \\

The \textbf{element-pattern} (\blue{elPat}) allows the description of an anti-pattern which examines all elements in the diagram. During the analysis all elements $\N$ in the diagram $D$ are compared to the element-pattern. All elements which conform to the pattern are returned.
\begin{flalign}
\setcounter{equation}{0}
    & \bp \ \sem{ \red{(}\blue{elPat}_1  \red{\mymid}  \blue{elPat}_2 \red{)}}_{D}  = \sem{\blue{elPat}_1}_{D} \ \cup \ \sem{\blue{elPat}_2}_{D} \\
    & \bp \ \sem{\red{Element}}_{D} = \{ (n, \{n\}) \mymid n\in \N \} \\
    & \bp \ \sem{\red{Element} \ \blue{typeFil}}_{D} = \{ (n, \{n\}) \mymid n\in \N \wedge  (n,\_) \in \sem{\blue{typeFil}}_{\N} \} \\
    & \bp \ \sem{\red{Element} \ \red{\{} \blue{elPatFil} \red{\}}}_{D} = \\
    &\hspace{0.5cm} \{ (n, \{n\} \cup M) \mymid n\in \N \wedge (n,M) \in \sem{\blue{elPatFil}}_{D,n}\} \\
    & \bp \ \sem{\red{Element} \ \blue{typeFil} \  \red{\{} \blue{elPatFil} \red{\}}}_{D} = \\
    & \hspace{0.5cm} \{ (n, \{n\} \cup M) \mymid n\in \N \wedge  (n,\_) \in \sem{\blue{typeFil}}_{\N}  \wedge \\
     & \hspace{0.7cm} (n,M) \in \sem{\blue{elPatFil}}_{D,n} \} 
\end{flalign} 
The first line (1) of every pattern term describes that the same pattern can occur several times if it is linked by a logical OR (\red{\mymid}). The evaluation of each pattern returns a set of tuples or an empty set. If all evaluations return an empty set, the empty set is returned. If one of the evaluations returns a non-empty set, this set is returned. If multiple evaluations return a non-empty set, the union of the sets is returned as each pattern can independently realize the threat.  \\
Line (2) displays the simplest form of the element-pattern, which does not specify any type or additional filters. Thus, this rule matches all elements and returns a set of tuples, whereby the number of tuples corresponds to the number of elements in the diagram since each tuple contains one of the elements. There are no additional affected components, therefore $\{n\}$ is returned. \\
(3) The resulting set contains only tuples whose elements correspond to the specified type-filter (\blue{typeFil}). The evaluation of the type-filter takes place under the consideration of the context capital $C$, which contains all element identifiers contained in $D$. \\
(4-5) This element-pattern includes one or more additional filters (\blue{elPatFil}). The context of the filter evaluation is the element identifier $n$. Some of the element-pattern-filters can affect additional diagram components which are contained in $M$. By design $\sem{\blue{elPatFil}}_{D,n}$ returns $\{(n,M)\}$ if the element $n$ passes the filter. If $n$ does not pass the filter, the empty set will be returned. \\
(6-8) Returns only elements that match the specified type-filter (\blue{typeFil}) and satisfy the additional filters. Similar to the lines (4-5), the resulting tuples contain the affected elements $M$.  \\

 The \textbf{asset-pattern} (\blue{assetPat}) allows the description of an anti-pattern which examines all assets inside the diagram. All assets $\Y$ inside the diagram $D$ are compared to the defined pattern. The evaluation is similar to that of the element-pattern (\blue{elPat}).
\begin{align*}
\setcounter{equation}{0}
    & \bp \ \sem{ \red{(}\blue{assetPat}_1  \red{\mymid}  \blue{assetPat}_2 \red{)}}_{D}  = \sem{\blue{assetPat}_1}_{D} \ \cup \ \sem{\blue{assetPat}_2}_{D} \\
    & \bp \ \sem{\red{Asset}}_{D} = \{ (y, \{y\}) \mymid y\in \Y \} \\
    & \bp \ \sem{\red{Asset} \ \blue{typeFil}}_{D} = \{ (y, \{y\}) \mymid y\in \Y \wedge  (y,\_) \in \sem{\blue{typeFil}}_{\Y} \} \\
    & \bp \ \sem{\red{Asset} \ \red{\{} \blue{assetPatFil} \red{\}}}_{D} = \\
    &\hspace{0.5cm} \{ (y, \{y\} \cup M) \mymid y\in \Y \wedge (y,M) \in \sem{\blue{assetPatFil}}_{D,y}\} \\
    & \bp \ \sem{\red{Asset} \ \blue{typeFil} \  \red{\{} \blue{assetPatFil} \red{\}}}_{D} = \\
    & \hspace{0.5cm} \{ (y, \{y\} \cup M) \mymid y\in \Y \wedge  (y,\_) \in \sem{\blue{typeFil}}_{\Y}  \wedge \ (y,M) \in \sem{\blue{assetPatFil}}_{D,y} \} 
\end{align*}

The  \textbf{boundary-pattern} (\blue{boundPat}) allows the description of an anti-pattern which examines all boundaries inside the diagram. All boundaries $\A$ inside the diagram $D$ are compared to the defined pattern. The evaluation is similar to that of the element-pattern (\blue{elPat}).

\begin{align*}
\setcounter{equation}{0}
    & \bp \ \sem{ \red{(}\blue{boundPat}_1  \red{\mymid}  \blue{boundPat}_2 \red{)}}_{D}  = \sem{\blue{boundPat}_1}_{D} \ \cup \ \sem{\blue{boundPat}_2}_{D} \\
    & \bp \ \sem{\red{Boundary}}_{D} = \{ (a, \{a\}) \mymid a\in \A \} \\
    & \bp \ \sem{\red{Boundary} \ \blue{typeFil}}_{D} = \{ (a, \{a\}) \mymid a\in \A \wedge  (a,\_) \in \sem{\blue{typeFil}}_{\A} \} \\
    & \bp \ \sem{\red{Boundary} \ \red{\{} \blue{boundPatFil} \red{\}}}_{D} = \\
    &\hspace{0.5cm} \{ (a, \{a\} \cup M) \mymid a\in \A \wedge (a,M) \in \sem{\blue{boundPatFil}}_{D,a}\} \\
    & \bp \ \sem{\red{Boundary} \ \blue{typeFil} \  \red{\{} \blue{boundPatFil} \red{\}}}_{D} = \\
    & \hspace{0.5cm} \{ (a, \{a\} \cup M) \mymid a\in \A \wedge  (a,\_) \in \sem{\blue{typeFil}}_{\A}  \wedge \  (a,M) \in \sem{\blue{boundPatFil}}_{D,a} \} 
\end{align*}

 The \textbf{connector-pattern} (\blue{conPat}) allows the description of an anti-pattern which examines all connectors inside the diagram. All connectors $\R$ inside the diagram $D$ are compared to the defined pattern.
 
\begin{flalign}
\setcounter{equation}{0}
    & \bp \ \sem{ \red{(}\blue{conPat}_1  \red{\mymid}  \blue{conPat}_2 \red{)}}_{D}  = \sem{\blue{conPat}_1}_{D} \ \cup \ \sem{\blue{conPat}_2}_{D} \\
    & \bp \ \sem{\red{Connector} \ (\red{\{} \blue{srcFil} \ \red{\&} \ \blue{tgtFil} \red{\}}}_{D} = \\
    & \hspace{0.5cm} \{ (r, \{r\} \cup M_1 \cup M_2)  \mymid r \in \R \\
	& \hspace{0.7cm} \wedge \exists n_1, (n_1, M_1) \in \sem{\blue{srcFil}}_{D}	\wedge \exists n_2, (n_2, M_2) \in \sem{\blue{tgtFil}}_{D} \\
	& \hspace{0.9cm} \wedge  source(r) = n_1 \wedge target(r) = n_2 \} \\
	& \bp \ \sem{\red{Connector} \ \blue{typeFil} \ (\red{\{} \blue{srcFil} \ \red{\&} \ \blue{tgtFil} \red{\}}}_{D} = \\
    & \hspace{0.5cm} \{ (r, \{r\} \cup M_1 \cup M_2)  \mymid r \in \R \wedge (r,\_) \ \in \ \sem{\blue{typeFil}}_{\R} \\
	& \hspace{0.7cm} \wedge \exists n_1, (n_1, M_1) \in \sem{\blue{srcFil}}_{D} 	\wedge \exists n_2, (n_2, M_2) \in \sem{\blue{tgtFil}}_{D} \\
	& \hspace{0.9cm} \wedge  source(r) = n_1 \wedge target(r) = n_2 \} \\
	& \bp \ \sem{\red{Connector} \ \blue{typeFil} \ (\red{\{} \blue{srcFil} \ \red{\&} \ \blue{tgtFil} \ \red{\&} \ \blue{conPatFil} \red{\}}}_{D} = \\
    & \hspace{0.5cm} \{ (r, \{r\} \cup M_1 \cup M_2 \cup M_3)  \mymid r \in \R \wedge (r,\_) \ \in \ \sem{\blue{typeFil}}_{\R} \\
	& \hspace{0.7cm} \wedge \exists n_1, (n_1, M_1) \in \sem{\blue{srcFil}}_{D} 	\wedge \exists n_2, (n_2, M_2) \in \sem{\blue{tgtFil}}_{D} \\
	& \hspace{0.9cm} \wedge  source(r) = n_1 \wedge target(r) = n_2 \\
	& \hspace{1.1cm} \wedge  (r,M_3) \in \sem{\blue{conPatFil}}_{D,r} \}
\end{flalign}
Lines (2-5) describe a connector-pattern without specified type-filter and no additional filters except the mandatory source and target-filter (\blue{srcFil},\blue{tgtFil}). Each resulting tuple consists of the connector $r$, the union of $r$, $M_1$ and $M_2$, where $M_1$ is a set that contains the source element $n_1$, and its affected components and $M_2$ is a set that contains the target element $n_2$ and its affected components.  \\
Lines (6-9) describe a connector-pattern variant with a specified type-filter. In addition to the lines (2-5), the connector identifier $r$ has to be inside the resulting set of the type-filter (\blue{typeFilter}) evaluation. \\
Lines (10-14) depict a connector-pattern variant with additional connector filters. The context of the filter evaluation is the connector identifier $r$. The additional affected components are contained in set $M_3$, which is also part of the union of each resulting tuple.  \\

The \textbf{flow-pattern} \blue{flowPat} allows the description of an anti-pattern which examines a defined flow sequence. In section \ref{sec:flow_definition}, the concept of flows has been introduced. The function $\mathsf{flows}(D,src,tgt)$ returns a set of all possible flow sequences between a source element and target element in diagram $D$.

\begin{flalign}
\setcounter{equation}{0}
    & \bp \ \sem{ \red{(}\blue{flowPat}_1  \red{\mymid}  \blue{flowPat}_2 \red{)}}_{D}  = \sem{\blue{flowPat}_1}_{D} \ \cup \ \sem{\blue{flowPat}_2}_{D} \\
    & \bp \ \sem{\red{Flow} \ \red{\{} \blue{srcFil} \ \red{\&} \ \blue{tgtFil} \red{\}}}_{D} = \\
    & \hspace{0.2cm} \{ (p, \{p\} \cup M_1 \cup M_2)  \mymid \exists n_1, (n_1, M_1) \in \sem{\blue{srcFil}}_{D} \\
	& \hspace{0.7cm} \wedge  \exists n_2, (n_2, M_2) \in \sem{\blue{tgtFil}}_{D} \wedge \exists p, p \in flows(D,n_1,n_2)\\
	& \hspace{0.9cm} \wedge  pSource(p) = n_1 \wedge pTarget(p) = n_2 \} \\
    & \bp \ \sem{\red{Flow} \ \red{\{} \blue{srcFil} \ \red{\&} \ \blue{tgtFil} \ \red{\&} \ \blue{flowPatFil} \red{\}}}_{D} = \\
    & \hspace{0.5cm} \{ (p, \{p\} \cup M_1 \cup M_2 \cup M_3)  \mymid \\
    & \hspace{0.7cm}\exists n_1, (n_1, M_1) \in \sem{\blue{srcFil}}_{D}   \\
	& \hspace{0.7cm} \wedge \exists n_2, (n_2, M_2) \in \sem{\blue{tgtFil}}_{D} \wedge \exists p, p \in flows(D,n_1,n_2) \\
	& \hspace{0.9cm} \wedge  pSource(p) = n_1 \wedge pTarget(p) = n_2 \\
	& \hspace{1.1cm} \wedge (p,M_3) \in \sem{\blue{flowPatFil}}_{G,p}\}
\end{flalign}
Lines (2-5) describe a flow-pattern with no additional filter, except the mandatory source and target-filter (\blue{srcFil},\blue{tgtFil}). Each resulting tuple consists of the flow sequence $p$ and the union of $p$, $M_1$ and $M_2$ where $M_1$ is a set that contains the source element $n_1$ and its affected components and $M_2$  is a set that contains the target element $n_2$ and its affected components.  The flow sequence $p$ must exist in the result set of the function $\mathsf{flows}(D,n_1,n_2)$ where $n_1$ is the source element and $n_2$ is the target element. \\
(6-11) This flow-pattern has additional flow-filters. The context of the filter evaluation is the flow sequence $p$. The additional affected components are contained in the set $M_3$, which is also part of the union of each resulting tuple. \\

The type filter (\blue{typeFil}) is used to examine the diagram components according to their assigned type. This filter can be applied to the element-pattern (\blue{elPat}), asset-pattern (\blue{assetPat}), boundary-pattern (\blue{boundPat}) and connector-pattern (\blue{conPat}). The evaluation of this filter takes place under the consideration of the special context capital $C$. This context contains the respective set of diagram component identifiers to which this filter is to be applied according to the previous pattern. 
\begin{flalign}
\setcounter{equation}{0}
    & \bp \ \sem{\red{: "} q \red{"} }_{C} =  \{ (c, \emptyset) \mymid  \lambda_C(c) \ = \ q  \ \vee \ \lambda_C(c) \ \in \ \iota_C(q)  \} \\
    & \bp \ \sem{\red{!= "} q \red{"} }_{C} =  \{ (c, \emptyset) \mymid  \lambda_C(c) \ \neq \ q \ \wedge \ \lambda_C(c) \ \notin \ \iota_C(q)  \} \\
    & \bp \ \sem{\red{: IN ["} q \red{", } \ldots , \red{"} q_i \red{"]} }_{C}  = \\
    &  \hspace{0.5cm} \{ (c, \emptyset) \mymid  \lambda_C(c) \ \in \ \{ q \ldots q_i \}  \ \vee \ \lambda_C(c) \ \in \ \{ \iota_C(q) \ldots  \iota_C(q_i)  \} \} \\
    & \bp \ \sem{\red{: NOT IN ["} q \red{", } \ldots , \red{"} q_i \red{"]} }_{C}  = \\  &  \hspace{0.5cm} \{ (c, \emptyset) \mymid  \lambda_C(c) \ \notin \ \{ q \ldots q_i \}  \ \wedge \ \lambda_C(c) \ \notin \ \{ \iota_C(q) \ldots  \iota_C(q_i)  \} \} 
\end{flalign}
The context capital $C$ contains either all element, asset, boundary or connector identifiers. The functions $\lambda_C$ and $\iota_C$ consider the set of identifiers specified inside the context capital $C$. The variable $q$ (or variables $q \ldots q_i$) defines the specified type which should be examined. \\
(1) The first version of this filter checks whether the assigned type of the diagram component matches the specified type $q$. Due to the fact that the types are defined in a two level hierarchy and each diagram component has both a "Top-Type" and a "sub-type", the component must either have exactly ($\lambda_C(c) \red{=} q$) the specified type ($q$) or be a "Sub-Type" of it ($\lambda_C(c) \in \iota_C(q)$). The returned set of the evaluation consists of tuples that contain the diagram component $c$ that fulfills the type-filter and the empty-set $\emptyset$ since no additional diagram components can be affected by this filter.\\
Line (2) displays the negated version of the first version (1). Therefore, the component must not have ($\lambda_C(c) \ \neq \ q$) the specified type ($q$) AND is not allowed to be a "Sub-Type" of it ($\lambda_C(c) \ \notin \ \iota_C(q)$). \\
(3-4) The third version of the type-filter is similar to the first version (1). However, in this case it is examined whether the assigned type of diagram component is contained within a set of possible types ($q_1 \ldots q_i$). Therefore, the assigned type must be in the set ($\lambda_C(c) \ \in \ \{ q \ldots q_i \}$), OR it is in the set of the "Sub-Types" ($\lambda_C(c) \ \in \ \{ \iota_C(q) \ldots  \iota_C(q_i)  \}$). \\
The last version of this filter (5-6) is the negated form of version three (3-4). Furthermore, it is similar to the second version (2).  Therefore, it is defined that the tuples of the resulting set can not contain diagram components that have an assigned type that is included inside the type-filter ($\lambda_C(c) \ \notin \ \{ q \ldots q_i \}$), AND the type must not be in the set of the "Sub-Types" ($\lambda_C(c) \ \notin \ \{ \iota_C(q) \ldots  \iota_C(q_i)  \}$). \\

 The \textbf{source-filter}  (\blue{srcFil})  is used to define the source element of a connector or flow-pattern. The \textbf{target-filter} (\blue{tgtFil}) is used to define the target element.
\begin{flalign}
\setcounter{equation}{0}
    & \ \sem{\red{Source} \ \blue{elPat}}_{D} = \{ (n,M) \mymid n \in N,  (n,M) \in \sem{\blue{elPat}}_{D}\} \\
    & \ \sem{\red{Target} \ \blue{elPat}}_{D} = \{ (n,M) \mymid  n \in N,  (n,M) \in \sem{\blue{elPat}}_{D}\}
\end{flalign}
(1-2) The resulting set includes only tuples that contain an element $n$ that conforms the specified element-pattern (\blue{elPat}), a set $M$ that contains the additional affected components. This applies to the source-filter as well as the target-filter. \\

The \blue{elPatFil} term defines the additional filters of an element-pattern. Therefore, the context $c$ can only be an element identifier. Assignable filters are:  \blue{propFil}, \blue{assetFil}, \blue{elRelFil}, \blue{conFil} and \blue{flowFil}.
\begin{align}
\setcounter{equation}{0}
    & \bp \ \sem{\blue{elPatFil}_1 \ \red{\&} \ \blue{elPatFil}_2}_{D,c} = \\
    & \hspace{0.5cm}\{ (c, M_1 \cup M_2) \mymid (c, M_1) \in \sem{\blue{elPatFil}_1}_{D,c}  \wedge (c, M_2) \in \sem{\blue{elPatFil}_2}_{D,c} \} \\
    & \bp \ \sem{\red{(}\blue{elPatFil}_1 \ \red{\mymid} \ \blue{elPatFil}_2\red{)}}_{D,c} = \sem{\blue{elPatFil}_1}_{D,c} \ \cup \ \sem{\blue{elPatFil}_2}_{D,c}
\end{align}
(1-2) If a pattern has multiple additional filters that are linked by a logical AND (\red{\&}), each one of them must be fulfilled by the component in the context $c$.
If one of the terms returns an empty set, the end result is the empty set. If all terms return a non-empty set, the end result is the cross-product of all the results, meaning all combinations of affected elements ($M_1$ and $M_2$) are returned. \\
(3) If the filters are linked by a logical OR (\red{\mymid}), only one of them has to be met by the component in the context $c$. The resulting set is the union of all evaluation sets as each entry of the evaluations independently fulfills the filter.  \\

The \blue{assetPatFil} term defines the additional filters of an asset-pattern. Therefore, the context $c$ can only be an asset identifier. The only assignable filter is the property-filter (\blue{propFil}).
\begin{align}
\setcounter{equation}{0}
    & \bp \ \sem{\blue{assetPatFil}_1 \ \red{\&} \ \blue{assetPatFil}_2}_{D,c} = \\
    & \hspace{0.5cm}\{ (c, M_1 \cup M_2 ) \mymid (c, M_1) \in \sem{\blue{assetPatFil}_1}_{D,c} \  \wedge (c, M_2) \in \sem{\blue{assetPatFil}_2}_{D,c} \} \\
    & \bp \ \sem{\red{(}\blue{assetPatFil}_1 \ \red{\mymid} \ \blue{assetPatFil}_2\red{)}}_{D,c} = \  \sem{\blue{assetPatFil}_1}_{D,c} \ \cup \ \sem{\blue{assetPatFil}_2}_{D,c}
\end{align}

 The \blue{boundPatFil} term defines the additional filters of a boundary-pattern. Therefore, the context $c$ can only be a boundary identifier. The only assignable filter is the boundary-relation-filter \blue{boundRelFil}.
\begin{align}
\setcounter{equation}{0}
    & \bp \ \sem{\blue{boundPatFil}_1 \ \red{\&} \ \blue{boundPatFil}_2}_{D,c} = \\
    & \hspace{0.5cm}\{ (c, M_1 \cup M_2 ) \mymid (c, M_1) \in \sem{\blue{boundPatFil}_1}_{D,c} \  \wedge (c, M_2) \in \sem{\blue{boundPatFil}_2}_{D,c} \} \\
    & \bp \ \sem{\red{(}\blue{boundPatFil}_1 \ \red{\mymid} \ \blue{boundPatFil}_2\red{)}}_{D,c} = \  \sem{\blue{boundPatFil}_1}_{D,c} \ \cup \ \sem{\blue{boundPatFil}_2}_{D,c}
\end{align}

 The \blue{conPatFil} term defines the additional filters of a connector-pattern. Therefore, the context $c$ can only be a connector. Assignable filters are:  \blue{propFil}, \blue{assetFil}, \blue{conCrossesFil}.
\begin{align}
\setcounter{equation}{0}
    & \bp \ \sem{\blue{conPatFil}_1 \ \red{\&} \ \blue{conPatFil}_2}_{D,c} = \\
    & \hspace{0.5cm}\{ (c, M_1 \cup M_2 ) \mymid (c, M_1) \in \sem{\blue{conPatFil}_1}_{D,c} \  \wedge (c, M_2) \in \sem{\blue{conPatFil}_2}_{D,c} \} \\
    & \bp \ \sem{\red{(}\blue{conPatFil}_1 \ \red{\mymid} \ \blue{conPatFil}_2\red{)}}_{D,c} = \  \sem{\blue{conPatFil}_1}_{D,c} \ \cup \ \sem{\blue{conPatFil}_2}_{D,c}
\end{align}

The \blue{flowPatFil} term  defines the additional filters of a flow-pattern (\blue{flowPat}). Therefore, the context $c$ can only be a flow ($p$). Assignable filters are: \blue{flowCrossesFil}, \blue{includesFil}.

\begin{align}
\setcounter{equation}{0}
    & \bp \ \sem{\blue{flowPatFil}_1 \ \red{\&} \ \blue{flowPatFil}_2}_{D,c} = \\
    & \hspace{0.5cm}\{ (c, M_1 \cup M_2 ) \mymid (c, M_1) \in \sem{\blue{flowPatFil}_1}_{D,c} \  \wedge (c, M_2) \in \sem{\blue{flowPatFil}_2}_{D,c} \} \\
    & \bp \ \sem{\red{(}\blue{flowPatFil}_1 \ \red{\mymid} \ \blue{flowPatFil}_2\red{)}}_{D,c} = \ \sem{\blue{flowPatFil}_1}_{D,c} \ \cup \ \sem{\blue{flowPatFil}_2}_{D,c}
\end{align}

The \textbf{property-filter} (\blue{propFil}) investigates the assigned properties of the component given by the context $c$. The context can either be an element, an asset or a connector identifier. 

\begin{flalign}
\setcounter{equation}{0}
    & \bp \ \sem{\red{"} k \red{"}\ \red{=}\ \red{"} v \red{"}}_{D,c} = 
        \left \{ \begin{array}{lll}
			    \{(c, \{c\}) \}      & \text{if} \ \mu(c,k) \ \equiv \ v    \\
				\emptyset               &  \text{if} \ \mu(c,k) \ \not\equiv \ v  \vee \mu(c,k) \ =  \ \textit{undefined}
			\end{array} \right. \\
	& \bp \ \sem{\red{"} k \red{"}\ \red{!=}\ \red{"} v \red{"}}_{D,c} = 
        \left \{ \begin{array}{lll}
			    \{(c, \{c\}) \}      & \text{if} \ \mu(c,k) \ \not\equiv \ v     \\
				\emptyset               &  \text{if} \ \mu(c,k) \ \equiv \ v  \vee \mu(c,k) \ =  \ \textit{undefined}
			\end{array} \right. \\
	& \bp \ \sem{\red{"} k \red{"}\ \red{in}\ \red{[} \red{"} v_1 \red{"} \red{, } \ldots \red{"} v_i \red{"} \red{]}}_{D,c} =  \\
       & \hspace{1cm}  \left \{ \begin{array}{lll}
			\{(c, \{c\})\}  & \  \text{if} \ \mu(c,k) \ \in \ \{ v_1, \ldots v_i \}    \\
			\emptyset	        & \  \text{if} \ \mu(c,k) \ \notin \ \{ v_1, \ldots v_i \}  \vee \mu(c,k) \ =  \ \textit{undefined}
			\end{array} \right. \\
	& \bp \ \sem{\red{"} k \red{"}\ \red{in}\ \red{[} \red{"} v_1 \red{"} \red{, } \ldots \red{"} v_i \red{"} \red{]}}_{D,c} = \\
       & \hspace{1cm}
         \left \{ \begin{array}{lll}
			\{(c, \{c\})\}  & \ \text{if} \ \mu(c,k) \ \notin \ \{ v_1, \ldots v_i \}     \\
			\emptyset	        &  \ \text{if} \ \mu(c,k) \ \in \ \{ v_1, \ldots v_i \} \vee \mu(c,k) \ =  \ \textit{undefined}
			\end{array} \right. 
\end{flalign}
(1) Returns the provided context $c$  if the function $\mu(c,k)$ returns the same value as defined in $v$. If the result of the function $\mu(c,k)$  does not match the value $v$, the empty set is returned.  Since the function $\mu$ is a partial function, it is also possible that the property key $k$ is not defined for the component in context $c$ and is therefore \emph{undefined}. If $k$ is \emph{undefined} for the context $c$, the empty set is returned. \\
(2) Returns the provided context $c$  if the function $\mu(c,k)$ does not return the same value as defined in $v$ or the partial function is not defined for the property $k$ and the context $c$. An empty set is returned if the result of $\mu(c,k)$ matches the value $v$.\\
(3-4) Is similar to the first variant of the property-filter, except that in this case a set of possible values $(v_1 \ldots v_i)$ is specified. The context $c$ is returned if the result of the partial function $ \mu(c, k) $ is in the set of values or is \emph{undefined}. The empty set is returned if the result of the function $ \mu(c, k) $ is not in the values $(v_1 \ldots v_i)$.\\
(5-6) Represents the negated form of the variant from lines (3-4). In this case, the context $c$ is returned if the result of the partial function $ \mu(c, k) $ is not within the set of the specified values or it is \emph{undefined}. An empty set is returned if the result of $\mu(c,k)$ is part of the values set $v_1 \ldots v_i$.\\

 The \textbf{asset-filter} (\blue{assetFil}) examines whether an element or connector is linked to an asset defined by the asset-pattern (\blue{assetPat}). This filter can only be assigned to an element  or a connector-pattern. For this reason, context $c$ can only contain one element or connector identifier.
\begin{flalign}
\setcounter{equation}{0}
    & \bp \ \sem{\red{Holds } \blue{assetPat}}_{D,c} = \ \{ (c, M) \mymid \exists y,  (y,M) \in \sem{\blue{assetPat}}_{D} \wedge (c,y) \in \rho \} 
\end{flalign}
(1) The diagram $D$ contains asset $y$, which corresponds to the specified asset-pattern (\blue{assetPat}). The relation $\rho$ contains all relationships between elements, connectors and assets. Therefore, the tuple $(c, y)$ must be contained in $\rho$ for this filter to apply. The resulting set includes only tuples that contain the context $c$ that is related to the asset $y$, a set $M$ that contains the affected asset.   \\

 The \textbf{element-relation-filter} \blue{elRelFil} examines the relations of an element component with other elements and boundaries. This filter can only be applied to an element-pattern (\blue{elPat}). Therefore, the context $c$ can only be an element identifier $n$. 
\begin{flalign}
\setcounter{equation}{0}
    & \bp \ \sem{\red{Contains } \blue{elPat}}_{D,c} = \ \{ (c, M) \mymid \exists n,  (n,M) \in \sem{\blue{elPat}}_{D} \wedge (c,n) \in \delta \} \\
    & \bp \ \sem{\red{Contains no } \blue{elPat}}_{D,c} = \ \{ (c, \{c\}) \mymid  \sem{\red{Contains } \blue{elPat}}_{D,c} = \emptyset \} \\
    & \bp \ \sem{\red{Contained by } \blue{elPat}}_{D,c} = \  \{ (c,  M) \mymid  \exists n,  (n,M) \in \sem{\blue{elPat}}_{D} \wedge (n,c) \in \delta \} \\
    & \bp \ \sem{\red{Not Contained by } \blue{elPat}}_{D,c} = \  \{ (c, \{c\}) \mymid  \sem{\red{Contained By } \blue{elPat}}_{D,c} = \emptyset \} \\
    & \bp \ \sem{\red{Contained by } \blue{boundPat}}_{D,c} = \  \{ (c, M) \mymid  \exists a,  (a,M) \in \sem{\blue{boundPat}}_{D} \wedge (a,c) \in \kappa \} \\
    & \bp \ \sem{\red{Not Contained by } \blue{boundPat}}_{D,c} = \  \{ (c, \{c\}) \mymid  \sem{\red{Contained By } \blue{boundPat}}_{D,c} = \emptyset \}
\end{flalign} 
(1) Examines if the element in context $c$ \red{contains} another element $n$. This element has to conform to the defined element-pattern (\blue{elPat}), and the tuple $(c,n)$ must be part of the relation $\delta$ which defines all element-element relations. The resulting tuples contain the context $c$, the set $M$ which contains the element $n$, and the additional affected diagram components. \\
(2) Represents the negated form of (1). Therefore, the resulting set of (1) has to be the empty set.  As a non-existing element cannot be part of the result, only the element identifier in context $c$ is returned. \\
(3) Examines if the element in context $c$ is \red{contained by} another element $n$. This element has to conform the defined element-pattern (\blue{elPat}) and the tuple $(n,c)$ must be part of the relation $\delta$. \\
(4) Represents the negated form of (3). Therefore, the resulting set of (3) is not allowed to contain a tuple which includes the context $c$.\\
(5) Checks if the element in context $c$ is \red{contained by} a boundary $a$. This boundary has to conform to the defined boundary-pattern (\blue{boundPat}), and the tuple $(a,c)$ must be part of the relation $\kappa$ which defines all boundary-element relations. The resulting tuples contain the context $c$, the set $M$ which contains the boundary $a$, and the additional affected diagram components. \\
(6) Represents the negated form of (5). Therefore, the resulting set of (5) has to be the empty set.\\

 The \textbf{boundary-relation-filter} \blue{boundRelFil} examines the relations of an boundary component with other elements and boundaries. This filter can only be applied to an boundary-pattern (\blue{boundPat}). Therefore, the context $c$ can only be a boundary identifier $a$. 
\begin{flalign}
\setcounter{equation}{0}
    & \bp \ \sem{\red{Contains } \blue{elPat}}_{D,c} =  \  \{ (c, M ) \mymid \exists n,  (n,M) \in \sem{\blue{elPat}}_{D} \wedge (c,n) \in \kappa \} \\
    & \bp \ \sem{\red{Contains no } \blue{elPat}}_{D,c} = \ \{ (c, \{c\}) \mymid \sem{\red{Contains } \blue{elPat}}_{D,c} = \emptyset \} \\
    & \bp \ \sem{\red{Contains } \blue{boundPat}}_{D,c} = \ \{ (c, M) \mymid \exists a,  (a,M) \in \sem{\blue{boundPat}}_{D} \wedge (c,a) \in \kappa \} \\
    & \bp \ \sem{\red{Contains no } \blue{boundPat}}_{D,c} = \  \{ (c, \{c\}) \mymid \sem{\red{Contains } \blue{boundPat}}_{D,c}  = \emptyset \} \\
     & \bp \ \sem{\red{Contained by } \blue{boundPat}}_{D,c} = \  \{ (c, M) \mymid  \exists a,  (a,M) \in \sem{\blue{boundPat}}_{D} \wedge (a,c) \in \kappa \} \\
    & \bp \ \sem{\red{Not Contained by } \blue{boundPat}}_{D,c} = \  \{ (c, \{c\}) \mymid \sem{\red{Contained by } \blue{boundPat}}_{D,c}  = \emptyset \} 
\end{flalign}
(1-6) The evaluation of this filter is similar to the previously presented element-relation-filter (\blue{elRelFil}), except that the relationships of a boundary are examined.\\

\clearpage

 The \textbf{connector-filter} (\blue{conFil}) examines the incoming and outgoing connections of an element. This filter can only be applied to an element-pattern (\blue{elPat}). Therefore, the context $c$ can only be an element identifier $n$.
\begin{flalign}
\setcounter{equation}{0}
    & \bp \ \sem{\red{Has Connector}\ (\blue{typeFil})? \ \red{\{} \blue{srcFil}\ \red{\&} \ \blue{conPatFil} \red{\}} }_{D,c} = \\
    & \hspace{0.4cm} \{(c,\{r \} \cup M_1 \cup M_2) \mymid r \in \R \wedge \exists n_1, (n_1, M_1) \in \sem{\blue{srcFil}}_{D} \\
    & \hspace{0.7cm} \wedge (\blue{typeFil} = \nv \vee (r,\_) \in \blue{typeFil}_{\T} ) \\
    & \hspace{0.9cm} \wedge  source(r) = n_1 \wedge target(r) = c  \ \wedge (r,M_2) \in \sem{\blue{conPatFil}}_{D,r} \} \\
    & \bp \ \sem{\red{Has Connector}\ (\blue{typeFil})? \ \red{\{} \blue{tgtFil}\ \red{\&} \ \blue{conPatFil} \red{\}} }_{D,c} = \\
    & \hspace{0.4cm} \{(c,\{r\} \cup M_1 \cup M_2) \mymid  r \in \R \wedge \exists n_1, (n_1, M_1) \in \sem{\blue{tgtFil}}_{D} \\
    & \hspace{0.7cm} \wedge (\blue{typeFil} = \nv \vee (r,\_) \in \blue{typeFil}_{\T} ) \\
    & \hspace{0.9cm} \wedge  source(r) = c \wedge target(r) = n_1   \wedge (r,M_2) \in \sem{\blue{conPatFil}}_{D,r} \} \\
    & \bp \ \sem{\red{Has No Connector}\ (\blue{typeFil})? \ \red{\{} \blue{srcFil}\ \red{\&} \ \blue{conPatFil} \red{\}} }_{D,c} = \\
    & \hspace{0.3cm} \{(c,\{c\}) \mymid  \sem{\red{Has Connector}\ (\blue{typeFil})? \ \red{\{} \blue{srcFil}\ \red{\&} \ \blue{conPatFil} \red{\}} }_{D,c} = \emptyset \} \\
     & \bp \ \sem{\red{Has No Connector}\ (\blue{typeFil})? \ \red{\{} \blue{tgtFil}\ \red{\&} \ \blue{conPatFil} \red{\}} }_{D,c} = \\
    & \hspace{0.3cm} \{(c,\{c\}) \mymid \sem{\red{Has Connector}\ (\blue{typeFil})? \ \red{\{} \blue{tgtFil}\ \red{\&} \ \blue{conPatFil} \red{\}} }_{D,c} = \emptyset \} 
\end{flalign}
(1-4) Examines an incoming connector $r$ since the connector-filter contains a source-filter (\blue{srcFil}). Each result tuple consists of the context $c$ and the union of $r$, $M_1$ and $M_2$ where $M_1$ is a set that contains the source element $n_1$ and the affected diagram components. The additional affected components of the connector-filter (\blue{conPatFil}) are contained in the set $M_2$.   \\
Lines (5-8) are similar to (1-4), except that in this case, an outgoing connector is evaluated since a target-filter (\blue{tgtFil}) is specified. \\
(9-10) Represents the negated form of (1-4). Therefore, the resulting set of (1-4) has to be the empty set. As a non-existing connector cannot be part of the result, only the element identifier in context $c$ is returned. \\
(11-12) Represents the negated form of (5-8). \\

\clearpage

 The \textbf{flow-filter} (\blue{flowFil}) examines the incoming and outgoing flows of an element. This filter can only be applied to an element-pattern (\blue{elPat}). Therefore, the context $c$ can only be an element identifier $n$.
\begin{flalign}
\setcounter{equation}{0}
        & \bp \ \sem{\red{Has Flow}\  \red{\{} \blue{srcFil}\ \red{\&} \ \blue{flowPatFil} \red{\}} }_{D,c} = \\
    & \hspace{0.5cm} \{(c,\{p\} \cup M_1 \cup M_2) \mymid  \wedge \exists n_1, (n_1,M_1) \in \sem{\blue{srcFil}}_D \\
    & \hspace{0.7cm}  \wedge \exists p, p \in flows(D,n_1,c) \wedge (p,M_2) \in \sem{\blue{flowPatFil}}_{D,p} \} \\
    & \bp \ \sem{\red{Has Flow}\  \red{\{} \blue{tgtFil}\ \red{\&} \ \blue{flowPatFil} \red{\}} }_{D,c} = \\
    & \hspace{0.5cm} \{(c,\{p\} \cup M_1 \cup M_2) \mymid  \wedge \exists n_1, (n_1,M_1) \in \sem{\blue{tgtFil}}_D \\
    & \hspace{0.7cm}  \wedge \exists p, p \in flows(D,c,n_2) \wedge (p,M_2) \in \sem{\blue{flowPatFil}}_{D,p} \} \\
    & \bp \ \sem{\red{Has No Flow}\  \red{\{} \blue{srcFil}\ \red{\&} \ \blue{flowPatFil} \red{\}} }_{D,c} = \\
    & \hspace{0.5cm} \{(c,\{c\}) \mymid \sem{\red{Has Flow}\  \red{\{} \blue{srcFil}\ \red{\&} \ \blue{flowPatFil} \red{\}} }_{D,c}  = \emptyset\} \\
     & \bp \ \sem{\red{Has No Flow}\  \red{\{} \blue{tgtFil}\ \red{\&} \ \blue{flowPatFil} \red{\}} }_{D,c} = \\
    & \hspace{0.5cm} \{(c,\{c\}) \mymid \sem{\red{Has Flow}\  \red{\{} \blue{tgtFil}\ \red{\&} \ \blue{flowPatFil} \red{\}} }_{D,c} = \emptyset \} 
\end{flalign}
Lines (1-3) describe a flow-filter, which examines an incoming flow sequence $p$, as the flow-filter contains a source-filter (\blue{srcFil}). The function $\mathsf{flows}(D,n_1,c)$ returns a set of flows between the source element $n_1$ and the target element in the context $c$.  Each result tuple consists of the context $c$ and the union of $p$, $M_1$ and $M_2$ where $M_1$ is a set that contains the source element $n_1$ and the affected diagram components. The additional affected components of the flow-filter (\blue{flowPatFil}) are contained in the set $M_2$.   \\
Lines (4-6) are similar to (1-4), except that in this case, an outgoing flow is evaluated since a target-filter (\blue{tgtFil}) is specified. The function $\mathsf{flows}(D,c,n_2)$ returns a set of flows between the source element in the context $c$ and the target element $n_2$.  \\
Lines (7-8) represent the negated form of (1-4). Therefore, the resulting set of (1-4) has to be the empty set. As a non-existing flow sequence cannot be part of the result, only the element identifier in context $c$ is returned. \\
Lines (9-10) represent the negated form of (5-8). \\

\clearpage

 The \textbf{connector-crosses-filter} (\blue{conCrossesFil}) examines whether a connector crosses a boundary or an element. This filter can only be applied to a connector-pattern (\blue{conPat}) or a connector-filter (\blue{conFil}). Therefore, the context $c$ can only be a connector identifier $r$.
\begin{flalign}
\setcounter{equation}{0}
    & \bp \ \sem{\red{Crosses} \ \blue{elPat}}_{D,c} = \\
    & \hspace{0.5cm} \{(c, M) \mymid \exists n, (n,M) \in \sem{\blue{elPat}}_{D} \\
    & \hspace{0.7cm} \wedge (n,source(c)) \in \delta \neq  (n,target(c)) \in \delta  \\
    & \bp \ \sem{\red{Crosses} \ \blue{boundPat}}_{D,c} = \\
    & \hspace{0.5cm} \{(c, M) \mymid \exists a, (a,M) \in \sem{\blue{boundPat}}_{D} \\
    & \hspace{0.7cm} \wedge (a,source(c)) \in \kappa \neq  (a,target(c) \in \kappa) ) 
\end{flalign}
Lines (1-3) represent the first variant of this filter, which examines whether the connector in the context $c$ crosses an element $n$. The element $n$ has to conform to the specified element-pattern (\blue{elPat}). The relation $\delta$ contains all element relations inside the diagram $D$. A connector $r$ crosses the element if either the source element or the target element of the connector is contained by $n$, and the other is not contained by $n$. The resulting tuples contain the context $c$, a set $M$ which contains the crossed element, the additional affected elements.\\
Lines (4-6) depict the second variant of this filter, which examines whether the connector in the context $c$ crosses a boundary $a$. The boundary $a$ has to conform to the specified boundary-pattern (\blue{boundPat}). The relation $\kappa$ contains all boundary-element relations inside the diagram $D$. A connector $r$ crosses the element if either the source element or the target element of the connector is contained by $a$, and the other is not contained by $a$.\\

The \textbf{flow-crosses-filter} (\blue{flowCrossesFil}) examines whether any connector in the flow sequence crosses a boundary or an element. This filter can only be applied to a flow-pattern (\blue{flowPat}) or a flow-filter (\blue{flowFil}). Therefore, the context $c$ can only be a flow sequence $p$.
\begin{flalign}
\setcounter{equation}{0}
    & \bp \ \sem{\red{Crosses} \ \blue{elPat}}_{D,c} = \\
    & \hspace{0.5cm} \{(c,M) \mymid \exists n, (n,M) \in \sem{\blue{elPat}}_{D} \wedge \exists r \in connectors(c) \\
    & \hspace{1cm} \wedge \ (n,source(r)) \in \delta \neq  (n,target(r)) \in \delta \\
    & \bp \ \sem{\red{Crosses} \ \blue{boundPat}}_{D,c} = \\
    & \hspace{0.5cm} \{(c,M) \mymid \exists a, (a,M) \in \sem{\blue{boundPat}}_{D} \wedge \exists r \in connectors(c) \\
    & \hspace{1cm} \wedge (a,source(r)) \in \kappa \neq  (a,target(r)) \in \kappa\} 
\end{flalign}
A flow can consist of several connectors. The flow-crosses-filter is used to check whether any of the connectors crosses a boundary or element. The function $ connectors(p) $ returns a set of connector identifiers, which contains all connectors in the flow sequence $p$.\\

\noindent Lines (1-4) represent the first variant of this filter, which examines whether any connector $r$ in the flow sequence $p$ crosses an element $n$. The element $n$ has to conform to the specified element-pattern (\blue{elPat}). The evaluation of whether a connector crosses an element is similar to the connector-crosses-filter.\\
(5-8) The second variant of this filter examines whether any connector $r$ in the flow sequence $p$ crosses a boundary $a$. The boundary $a$ has to conform to the specified boundary-pattern (\blue{boundPat}). The evaluation of whether a connector crosses a boundary is similar to the connector-crosses-filter.\\

 The \textbf{includes-filter} (\blue{includesFil}) examines the elements or connectors in a flow sequence. This filter can only be applied to a flow-pattern (\blue{flowPat}) or a flow-filter (\blue{flowFil}). Therefore, the context $c$ can only be a flow sequence $p$.
\begin{flalign}
\setcounter{equation}{0}
  & \bp \ \sem{\red{Includes} \ \blue{elPat}}_{D,c} = \\
    & \hspace{0.5cm} \{(c, M) \mymid \exists n, (n,M) \in \sem{\blue{elPat}}_{D} \wedge n \in elements(c) \} \\
    & \bp \ \sem{\red{Includes no} \ \blue{elPat}}_{D,c} =  \{(c, \{c\}) \mymid \sem{\red{Includes} \ \blue{elPat}}_{D,c} = \emptyset \} \\
    & \bp \ \sem{\red{Includes only} \ \blue{elPat}}_{D,c} = \\
    & \hspace{0.5cm} \{(c,  M_1 \cup \ldots M_i) \mymid \{n_1, \ldots n_i\} = elements(c), \\
    & \hspace{1cm} (n_j,M_j) \in \sem{\blue{elPat}}_{D}, \ \forall 1 \leq j \leq  i \ \} \\
    & \bp \ \sem{\red{Includes} \ \blue{conPat}}_{D,c} = \\
    & \hspace{0.5cm} \{(c, M) \mymid \exists r, (r,M) \in \sem{\blue{conPat}}_{D} \wedge r \in connectors(c) \} \\
    & \bp \ \sem{\red{Includes no} \ \blue{conPat}}_{D,c} =  \{(c, \{c\}) \mymid \sem{\red{Includes} \ \blue{conPat}}_{D,c} = \emptyset \} \\
    & \bp \ \sem{\red{Includes only} \ \blue{conPat}}_{D,c} = \\
    & \hspace{0.5cm} \{(c, \{ M_1 \cup \ldots M_i\}) \mymid \{r_1, \ldots r_i\} = connectors(c), \\
    & \hspace{1cm} (r_j,M_j) \in \sem{\blue{conPat}}_{D}, \ \forall 1 \leq j \leq  i  \ \} 
\end{flalign}
Lines (1-2) represent the first variant of the includes-filter, which examines whether a flow sequence $p$ contains an element $n$. The element $n$ has to conform to the specified element-pattern (\blue{elPat}). The returned tuples contain the context $c$, a set $M$ that contains the additional affected components of the element-pattern.\\
(3) This filter variant represents the negated form of the first variant (1-2). Therefore, the resulting set of (1-2) has to be the empty set. As a non-existing element cannot be part of the result, only context $c$ is returned. \\
(4-6) The third variant of the filter examines whether all elements ($n_1 \ldots n_i$) in the a flow sequence $p$ conform to a specified element-pattern (\blue{elPat}).\\
Lines (7-8) depict an includes-filter which examines whether a flow sequence $p$ contains a connector $r$. The connector $r$ has to conform to the specified connector-pattern (\blue{conPat}). \\
Lines (9) represent the negated form of the filter in lines (8-9). \\
(10-12) The last filter variant examines whether all connectors ($r_1 \ldots r_i$) in the a flow sequence $p$ conform to a specified connector-pattern (\blue{conPat}).\\

\clearpage

\subsubsection{The Flow-finding Algorithm}
\label{sec:flows_algorithm}
\autoref{sec:flow_definition} introduced the concept of the flows. The function $ \mathsf{flows}(D, src, tgt) $ is used to find all possible flows between a source element $ src $ and a target element $ tgt $ within a diagram $ D $. The function is used within the flow-pattern (\blue{flowPat}) and the flow-filter (\blue{flowFil}). By definition, a flow consists of an alternating sequence of elements and connectors. The following algorithm describes the implementation of the function $ \mathsf{flows}(D, src, tgt) $. \\

\begin{algorithm}
\caption{The Flow-finding Algorithm}
\begin{algorithmic}[1]
\State --Initialization --
\State Input: Diagram (D)
\State Input: Source Element (src)
\State Input: Target Element (tgt)
\State Output: Set of Flows (flowSet)
\Procedure {flows}{$D$, $src$, $tgt$}
    \State  $flowSet \gets \emptyset$
    \State  $usedConnectors \gets \emptyset$
    \State  \Call{recursiveSearch}{$src$,$tgt$,$usedConnectors$,$flowSet$,$D$}
    \State  \textbf{return} $flowSet$
\EndProcedure
\Statex
\Procedure {recursiveSearch}{$src,tgt,usedConnectors,flowSet$}
    \If{$|usedConnectors| > 0$}
        \If{$src \equiv tgt$}
            \State $p \gets createSeq(usedConnectors)$
            \State  $flowset \ \textbf{add} \ p$
            \State  \textbf{return}
        \EndIf
    \EndIf
    \State $srcConnectors \gets connectors(D,src)$
    \For{\textbf{each} $connector \in srcConnectors$}
         \If{$connector \notin usedConnectors$}
            \State $newSrc \gets target(connector)$
            \State $usedConnectors \ \textbf{add} \ connector$
            \State  \Call{recursiveSearch}{$newSrc$,$tgt$,$usedConnectors$,$flowSet$,$D$}
            \State  $usedConnectors \ \textbf{remove} \ connector$
         \EndIf
    \EndFor
\EndProcedure
\end{algorithmic}
\end{algorithm}

\clearpage
The algorithm implements the depth-first search (DFS) approach. The analysis of an element is paused as soon as the next element has been identified \shortcite[p.48]{mukhopadhyay_complexity_2016}. \\

The lines (1-5) describe the initialization of the function. As already mentioned, the flows function takes three arguments: a diagram instance ($D$), a source element ($src$), and a target element ($tgt$). The function returns a set of flows as output. \\
Lines (6-11) describe the initial call of the function $\mathsf{flows}(D,src,tgt)$. Line (7) initializes a new $flowSet$ is as an empty set. (8) A flow is an alternating sequence of elements and connectors. However, it is defined that each connector in a flow sequence must be unique. Therefore, the algorithm has to remember which connectors have already been visited in order to avoid endless loops. $usedConnectors$ is an initially empty set that stores the visited connectors. Line (9) shows the call of the function $\mathsf{recursiveSearch}$. This function takes five parameters as input: source element,  target element, the set of already visited connectors ($usedConnectors$), the $flowSet$, and the diagram instance $D$. The arguments are passed as references to the function. (10) The $flowSet$ is the return/output value of the algorithm as it contains the end result of the algorithm. \\ 
The recursive function $recursiveSearch$ is depicted in lines (12-29). (13) If the function is called the first time, the set of the $usedConnectors$ is empty. A flow must contain at least one connector, and therefore, the number of visited connectors ($usedConnectors$)  must be greater than zero.\\
The lines (14-18) depict the end of a recursive call if the source element is equivalent to the target element. (15) If the source element is equivalent to the target element, a flow has been found. The set of $usedConnectors$ contains all connectors which are part of the flow $p$. The function $\mathsf{createSeq}(usedConnectors)$ converts the set of connectors to a valid flow sequence. (16) The flow sequence is added to the $flowSet$. Line (17) depicts the end of a recursive call. \\
(20) If the $usedConnectors$ set does not contain any connectors or the current source element does not match the target element, the algorithm must continue. The function $\mathsf{srcConnectors}(D,\allowbreak src)$ returns a set of connectors whose source element is the current $src$ element. These are stored inside the variable $srcConnectors$. \\
Lines (21-28) describe the search for the next source element for the next recursive iteration. (21-22) For each $connector$ within the $ srcConnectors $ set, it is checked whether it has already been visited. If not, then the target element of this connector is the new source element ($newSrc$) for the next iterative call. (24) The $connector$ is added to the set of $usedConnectors$. Line (25) depicts the call of the next recursive iteration. (26) The $connector$ is removed from the $usedConnectors$ as soon as the recursive call returns. \\ 
\section{Related Research/Tools}
\label{sec:related}
In this section, we look at the concept behind security analyses of IT-systems. To this end, we present the most common strategies, methods, existing tools and related research in this area. 

\subsection{Threat Modeling and Risk Management}
\label{sec:threat_modelin}

Before describing the analysis approach of ThreatGet, we introduce the concept of threat modeling. An adequate definition of the threat modeling concept at the high level is the following: \textit{"Threat modeling is the first step in any security solution. It's a way to start making sense of the vulnerability landscape. What are the real threats against the system? If you do not know that, how do you know what kind of countermeasures to employ?" } \cite[p.214]{schneier_threat_2000}. According to this quote by \citeauthor{schneier_threat_2000},  identifying vulnerabilities and threats using an adequate threat modeling methodology is essential to a successful risk management process \cite{schneier_threat_2000}. Consequently, threat modeling is a key subpoint in the larger process of risk management, as only threats and vulnerabilities that have already been identified can be addressed. \\

Risk management is not a straightforward process that ends after a single execution but should be performed iteratively to respond to new threats, information, and changes \cite{schmittner_threat_2019}.  To make the entire process as effective as possible, it is also important to involve as many stakeholders as possible and to take a holistic view of it. \\
Over time, three different threat modeling strategies have emerged. Most existing methods can be assigned to at least one of these strategies. However, there are also methods and frameworks that combine all strategic approaches, such as the "Hybrid Threat Modeling Method" (hTMM) by \citeA{mead_hybrid_2018}. In the following, we present the strategies and their representatives and show which of them has been implemented in ThreatGet. \\

\subsection{The Attacker-Centric Strategy } 
The focus of this strategy, as the name implies, is on the person behind the attack. The goal of this strategy is to identify the attacker archetype \shortcite{shevchenko_threat_2016}. Therefore, the analyst tries to put himself in the position of the attacker and thus simulate the activities. For instance, the motivation, the background and the necessary knowledge for an attack are analyzed.  A well-known method of this strategy is Persona-non-Grata (PnG). The main process of this method is to collect the information and document it on index cards. For this strategy, the approach and information extraction are difficult to automate because they are based on rather subjective and non-technical procedures. Nevertheless, this approach is used to get a first impression or to initiate the threat modeling process in general \cite{mead_hybrid_2018}.\\

\subsection{The Asset-Centric Strategy} 
The focus is on the assets of an organization or system. In broad terms, an asset denotes an object of interest or value.  The initial process of an asset-centric strategy is to identify all assets of a given system \citeauthor{eng_integrated_2017}.  It can be helpful to analyze what might be of interest to an attacker and, in particular, what is essential to the company or organization. Based on this, it is analyzed how an attack could damage the assets or how an aggressor could access the assets \shortcite{shostack_threat_2014}. \\
A well-known method is the CORAS framework, which defines a graphical modeling language for this purpose \shortcite{coras_2007}. A major drawback of this approach is that the set of assets can contain a large number of entries, depending on the number of stakeholders and the size of the system. In order to avoid scalability issues, the entries must be reviewed and ranked by importance. Subsequently, possible attacks and damage scenarios are worked out for all assets, starting with the most important ones. Similar to the attacker-centric strategy, most processes in this strategy are based on subjective rather than objective approaches. In particular, the identification and valuation of assets depends to a large extent on the stakeholders involved. In addition, the attack scenarios and damage scenarios are based on the experience of the analysts involved \shortcite{eng_integrated_2017, poller_asset_2014, shevchenko_threat_2016}. However, it makes sense to think about what data assets are in a system and how they could be compromised, as they can provide information about the impact and likelihood of an attack.  For this reason, ThreatGet allows the user to highlight assets in its diagram and consider them in the analysis.  \\ 

\subsection{The Software-Centric Strategy }
The primary focus of this approach is on models that represent the underlying architecture of a system and its operating software. The straightforward focus on system components facilitates involving stakeholders in the security analysis process. Designers and developers do not typically think in terms of business goals or assets. Therefore, it is easier to involve them in a process that relates directly to what they create \cite{shostack_threat_2014,gumbley_guide_2020}. Another positive aspect is that this approach can already be used during the planning phase of a project, as initial conceptual designs of the system can already be made there, which can be used for security analysis. Furthermore, existing systems can also be analyzed with the help of this process. \\

Different diagram types can be used to represent the systems, e.g., those from the standard "Unified Modeling Language" (UML) or "Data Flow Diagrams" (DFD). DFDs are often used because they have a simpler structure and are therefore easier to design. \autoref{sec:rel_dfd} takes a closer look at the data-flow diagram since most threat modeling methods are based on this diagram, and ThreatGet also uses an extended form of this diagram.. However, the decision for a diagram type should depend on the preferences of the developers.

The best-known threat modeling method that can be assigned to this strategy is STRIDE. This method was developed back in 1999 by Loren Kohnfelder, and Praerit Garg \cite{kohnfelder_1999}. The name STRIDE is an acronym composed of the first letters of the STRIDE categories: Spoofing, Tampering, Repudiation, Information Disclosure, Denial of Service, and Elevation of Privilege. The STRIDE method is intended as a mnemonic and threat categorization to assist the user in the security analysis of a system under consideration \cite{shostack_threat_2014}.\\

The method can be performed in two ways. The first approach is called "STRIDE-per-element" and focuses on each element within the system.  The analyst iterates over each element and examines it for all known threats within the STRIDE categories. This approach is very simple and can be performed by a relatively inexperienced person. \\
The second approach is called "STRIDE-per-Interaction" and focuses on the analysis of the data exchange between two components. The procedure here is similar to the first way, but in this case, more variables are involved in the process, as both the elements and the transmission path of data are analyzed. \textit{"STRIDE-per-element is a simplified approach to identifying threats, designed to be easily understood by the beginner. However, in reality, threats don't show up in a vacuum. They show up in the interactions of the system." } \cite[p.80]{shostack_threat_2014}.\\

\subsubsection{The Data-Flow Diagram}
\label{sec:rel_dfd}
Most of the tools that aim to automate security analysis, which we present in \autoref{sec:template_tools} and \autoref{sec:logic_tools}, are based on a STRIDE guided approach and a DFD. A data flow diagram is composed of a total of five different components, which are listed and specified below: 

\begin{itemize}
 \setlength\itemsep{0.2cm}
    \item \textbf{External Entity}: Describes a component that is not directly part of the system, but can interact with it. 
    
    \item \textbf{Data Store}: Represents some form of data storage. For example a database or file.
    
    \item \textbf{Process}: Describes an action that is executed within a system. 
    
    \item \textbf{Data-Flow}: Represents a potential data exchange between to components.
    
    \item \textbf{Boundary}: Describes a distinction between logically or physically separated system components. In addition, boundaries can be defined to represent multiple privilege levels within the system. They are not an actual part of the system, but can be modeled to represent an attack surface.\cite[p.49-50]{shostack_threat_2014}. 
\end{itemize}

Except for boundaries, properties can be assigned to all diagram components. Properties are defined in the form of key-value pairs and can provide additional information about the specific component. \\
By focusing on what is being developed and taking a structured approach to analyzing individual system components and their interaction possibilities, the software-centric strategy creates the best conditions for automating security analysis. \\
Due to this fact, several tools have already been developed that focus on assisting with analysis or implement an automated approach. Four of these tools have been studied prior to the development of ThreatGet, and we present them in the following two sections. \\
We divide these tools into two categories because three of each are very similar in their approach to analysis.

\subsection{Template-based Analysis Tools}
\label{sec:template_tools}
The first group of software-centric analysis tools includes two of the total five, and we have classified these as template-based analysis or supporting tools. Both tools use the data flow diagram presented above to represent the system under investigation. They provide an interface in which the system can be modeled and modified, using predefined diagram component templates. We have placed these tools in this group because they also show the user which threats could occur during the analysis. This requires the user to define all potential threats in a template-based format and enter them into the database. As soon as the analysis is started, the user must work manually through the list of threats according to the "STRIDE-per-Element" and "STRIDE-per-Flow" principles and decide which of the entries may or may not apply. \\
ThreatGet uses an approach comparable to the tools from the second group. However, these are also presented here as they can make a non-negligible contribution to improving the security posture of systems.\\

The first tool we want to highlight is \textbf{Threat Dragon}. It is an open-source threat modeling tool developed by the Open Web Application Security Project (OWASP)\footnote{\url{https://docs.threatdragon.org/}}. Its interface provides a non-modifiable list of diagram component templates that can be used to create a data flow diagram. Every template, except for the boundary, comes with its own but not extendable set of properties. Using the tool is relatively straightforward, as creating the diagram and defining threats is quite simple. In addition to the text-based definition of the threats, the user can add severity and mitigation information to each threat entry. During the analysis, the tool supports the user in selecting threats by proposing threat entries based on the diagram component and its assigned properties.\\ 

The second tool is called \textbf{IriusRisk}. It is a commercial tool designed and developed by Continuum Security\footnote{\url{https://iriusrisk.com/}}. This tool also uses a template-based approach but offers a much more comprehensive functions compared to Threat Dragon. In contrast to Threat Dragon, the user can freely define individual diagram components and use them for system-modeling.  Furthermore, it is also possible to extend the set of assigned properties of the template with user-defined properties. The IriusRisk threat template consists of a textual description of the threat and additional information, such as an assessment of exploitability and potential impact. The defined threat templates can be linked to the component templates. The underlying concept of this approach is that systems are typically built with similar components. Especially when an organization reuses the same components for different products, therefore, once defined, the components and their threats can be reused in multiple diagrams. IriusRisk presents the user with the associated threats for a specific component or data flow during the analysis. This reduces the time needed. However, the user still has to decide which of the threats apply manually. \\

\subsection{Logic Analysis Tools}
\label{sec:logic_tools}
The second group of software-centric analysis tools includes the remaining two tools. We classified them as logic-based analysis tools, as they implement additional functionality to determine whether a defined threat is present in a given system.  \\
For this purpose, all three use a specific analysis language with their own syntax and semantic, defined in a context-free grammar (CFG). Such an analytical language allows threats not only to be described in plain text, but also to be interpreted by a machine. The formulation of the threat in such a language is denoted in the following as an anti-pattern. An anti-pattern describes a state or condition in a system that can lead to a particular threat. \\
The analysis languages consist of a set of specified sentences. These sentences are each composed of individually defined symbols. The grammar of the language gives the order of the sentences and symbols. \textit{"A context-free grammar (CFG) consists of a set of production rules. Each rule describes how a non-terminal symbol can be "replaced" or "expanded" by a string that consists of non-terminal symbols and terminal symbols."} \cite[p.2]{casanova_syntactic_2016}. Whereby a terminal symbol represents a fixed part of the language and a non-terminal symbol must be replaced by the given production rule until a terminal symbol is reached. \\
During the analysis, the tools iterate over the defined threat catalog, interpret the anti-patterns and automatically check whether the declared threat is present in the system under investigation. Due to this approach, only those threats are displayed that actually apply to the system. This significantly reduces the time required for threat identification. The time saved can be used more effectively for the evaluation and mitigation of the identified threats. In addition, non-sophisticated security analysts can perform this procedure, learn from it, and react directly to identified threats. ThreatGet also uses this approach and extends the functionality of the tools we are highlighting in this section. \\

\subsubsection{pyTM}
\label{sec:pyTM}
The first tool we want to discuss in this group is \textbf{pyTM}.  It is open-source tool which is also created an maintained by OWASP\footnote{\url{https://owasp.org/www-project-pytm/}}. Compared to the other tools, pyTM differs significantly in its approach, as it does not provide a graphical diagram editor. The system-model, i.e. the diagram, is defined in the python programming language. Components, their internal connections, and properties are defined in individual lines of code and added to the diagram object. Based on this code, pyTM provides the functionality to generate a visual diagram. In addition to the data-flow diagram presented in a previous section, it is also possible to create diagrams from the UML specification. \\
The tool provides a predefined library of diagram components and properties, which the user can extend.  Furthermore, the tool contains a catalog of threats that can also be extended. To manage the threats, the tool uses the JavaScript Object Notation (JSON) data format. The user can not only describe individual threats textually but also define an assessment of the potential impact and likelihood. Furthermore, descriptions regarding severity, necessary preconditions and potential mitigation options can be stored. However, this data is not included in the automatic analysis, but the user is shown this information when a threat is detected. The most important entry, which contains the anti-pattern in pyTM, is called "condition".  \\
The condition of a threat states the logical semantics for the evaluated component. The user can provide additional description of required circumstances for this threat, such as whether a component has a particular property, whether it is in a boundary, or whether it is connected to another component. These individual anti-patterns can also be logically linked to each other. Thus it can be stated whether both conditions must take place or only one of them.\\
Unfortunately, we could not find a complete definition of the full analysis language, as well as its evaluation. The GitHub account\footnote{\url{https://github.com/izar/pytm/}} includes some predefined examples. Furthermore, pyTM is a purely console-based tool and does not provide a graphical user interface (GUI). Only the generated diagrams and the analysis result are presented to the user. Therefore, it is difficult to use for people without programming skills, especially during the diagram and threat definition.

\subsubsection{The Microsoft Threat Modeling Tool}
\label{sec:MTMT}
The Microsoft Threat Modelling Tool (MTMT) is probably the best-known threat modeling tool. It is developed as part of Microsoft's Security Development Lifecycle\footnote{\url{https://www.microsoft.com/en-us/securityengineering/sdl/threatmodeling}}. The tool is primarily used to implement the STRIDE-per-interaction approach. It includes a catalog of diagram components that can be used to develop a data-flow diagram in the tool's own modeling interface. The components, as well as the defined threat entries, are organized in templates that can be grouped for use in specific domains. 

\begin{figure}[htb]
    \centering
    \includegraphics[width=1\textwidth]{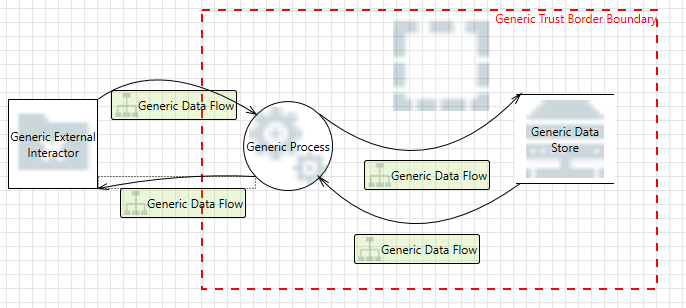}
    \caption{Microsoft Microsoft Threat Modeling Too Diagram}
    \label{fig:mtmt_example_diagram}
\end{figure}

\autoref{fig:mtmt_example_diagram} shows a rudimentary example diagram created using the MTMT. It shows three elements, each connected by data flows. This diagram is intended to represent the retrieval of data from a database by a cell phone. The left element, called the "Generic External Interactor," lies outside the "Generic Trust Boundary" and represents the cell phone. Inside the trust boundary lies a "Generic Process" that handles the query and the "Generic Data Store" containing the data. The flow could be as follows, the interaction sends a request to the process, which validates the request, retrieves the data, and returns it. The boundary represents the actual system, which can be accessed from outside. With this example, we want to show how easy it is to map such facts in a DFD. \\

To define threats with the MTMT, the user must define them in the analysis language provided by the tool. The grammar of this syntax is displayed in the following. The terminal symbols of the grammar are highlighted in red and non-terminals in blue.

\setcounter{equation}{0}
\begin{figure}[htb]%
  \fbox{%
    \parbox{\insidefboxwidth}{%
      \abovedisplayskip=0pt\belowdisplayskip=0pt%
      \begin{align*}
        &  \blue{rule}          & ::= & \ \blue{expression} \ ( \ \blue{op} \ \blue{expression}  \ )* \\
        & \blue{expression}    & ::= & \ \blue{object} \ ( \ \red{.[} \blue{literal} \red{]} \ )? \  \red{IS} \ \blue{value}\\
        &                            & & \mymid \red{Flow crosses} \ \blue{literal} 
                                    \mymid \red{NOT} \ \blue{expression} \\
        &  \blue{object}        & ::= & \ \red{Source} \mymid \red{Target} \mymid \red{Flow} \\
        &  \blue{value}         & ::= & \ \red{YES} \mymid  \red{NO} \mymid \blue{literal} \\ 
        &  \blue{op}            & ::= & \ \red{AND} \mymid  \red{OR} \\ 
        &  \blue{literal}       & ::= & \ \red{"}text\red{"} 
      \end{align*}%
    }%
  }
  \caption{Syntax of MTMT Language}
  \label{fig:MTMT_Language_Syntax}
\end{figure}

Further information to this grammar and its usage can be found in the Getting Started Guide \cite{microsoft_started_2016} and the User Guide \cite{microsoft_user_2016}. However, these documents provide no detailed description about the semantic evaluation during the analysis. Using this grammar, the following anti-patterns can be formulated, for example:

\begin{flalign}
\setcounter{equation}{0}
    & \bp \ \red{Source.["}  Secure Boot   \red{"] IS "} NO \red{"} \\
    & \bp \ \red{Target.["}  Authentication  \red{"] IS "} NO \red{"} \\
    & \bp \ \red{Source.["}  Secure Boot   \red{"] IS "} NO \red{"} \ \red{AND} \\
    & \hspace{0.4cm}\red{Target.["}  Authentication  \red{"] IS "} NO \red{"} \ \red{AND Flow crosses} \ \red{"} Boundary \red{"}
\end{flalign}
(1) The first anti-pattern checks whether some element that has an outgoing data-flow has the property "Secure Boot" set to "NO".\\
(2) The second anti-pattern checks whether some element that has an incoming data-flow has the property "Authentication" set to "NO".\\
(3-4) The last anti-pattern combines the previous ones and also checks, in addition, whether the data-flow between these two elements "crosses" a "boundary". In this example, the anti-pattern can be used to check if the mobile (Generic External Interactor) inside the boundary can communicate with the process without authenticating itself. This circumstance could mean that an attacker could send malicious data to our process and thus damage the system. A special feature of this analysis language is that it is possible to check whether a boundary is crossed, as these can also represent an attack surface. \\

The previous examples are intended to show how easy it is to create a system-model with the MTMT and analyze it using the anti-patterns. However, we have found a few limitations of the tool during testing. The MTMT only detects strict anti-patterns during analysis. That is, if the system model is not modeled exactly as described in the pattern, then the anti-pattern does not kick in. This fact is especially apparent when different people model the same system and then different results are displayed during analysis, even though the models are nearly indistinguishable. This drawback indicates that the modeling of the system and the definition of the threats are too closely related \cite{karahasanovic_adapting_2017}.

\clearpage 
\section{Conclusion}
\label{sec:conclusion}

IT-systems have changed significantly over the last two decades, and it is hard to imagine most economic sectors and most daily processes without them \shortcite{jang_jaccard_survey_2014,chen_great_2019}. The development towards increasingly interconnected devices is far from complete. It becomes even more apparent when we look at the current trends in the automotive sector and the Internet of Things. \\

In addition to the positive contributions and improvements that this development has brought, these connected systems are potentially more vulnerable to malicious attacks due to their openly accessible communication interfaces \cite{wolf_combining_2018}. This circumstance illustrates the importance of IT-Security. Nevertheless, many manufacturers neglect the security factor to save costs and compete on the free market \shortcite{gilchrist_iot_2017}. \\
Given the future mandatory security standards, the associated accountability, and the increasing awareness of customers regarding data security and integrity, these developers are under pressure to produce not only smart but also more secure devices. Since there will probably not be enough security experts to cover all domains and areas in the near future, there is a high demand for an automated solution to detect security vulnerabilities and threats. \\

For this reason, the Dependable Systems Engineering Group (DSE) at the Austrian Institute of Technology (AIT) has developed the security analysis tool ThreatGet. The tool allows users to automatically scan even complex IT-systems for potential vulnerabilities and threats in a consistent and repeatable way. \\
In this paper, we present the ThreatGet approach in detail. It is based on three essential components. The first component is an advanced data-flow diagram to represent a system under investigation and its security-related properties.  In \autoref{sec:diagram_example}, we present the modeling notation, as well as the complete formal definition of the diagram, using a basic example. \\
The second component is a threat and vulnerability knowledge database. In order to create and manage this threat knowledge base, a domain-specific analysis language has been developed that can be interpreted by both humans and machines. The language is used to define vulnerability and threat information in the form of anti-patterns. Each anti-pattern describes a potentially exploitable state or condition in the system that must be considered to create a secure product. The full syntax of the language is highlighted in \autoref{sec:rule_syntax}.\\
The third component is an automated analysis engine that takes the other components as input and compares them. The result of the analysis contains all the anti-patterns that are in the system. \autoref{sec:evaluation_patterns_filters} contains the full semantic evaluation of the anti-patterns, formulated in the analysis language. The semantic analysis allows each user to clearly understand how the analysis is performed and how the results are derived.  \\
This approach is similar to other system-centric approaches such as the Microsoft Threat Modeling Tool but extends existing solutions in all aspects.\\

ThreatGet is designed to improve the security posture of systems without the need for a security expert. However, it is not feasible to assume that one system can be totally secured. New security exploits are discovered on an almost daily basis. Nevertheless, a system has to be designed and developed to be as secure as possible. Therefore, the best approach to achieve this is to examine the system for known weaknesses and implement suitable security measures to defend it from potential threats. \\
The analysis approach of ThreatGet provides an automated approach towards vulnerability and threat identification.  It further extends the approaches of existing solutions. Furthermore, it features a persistent approach concerning the management of threat knowledge.

\clearpage 
\section{Future Work / Outlook}
\label{sec:future_work}
In this section, we discuss how we will extend the ThreatGet tool itself and the analysis language for formulating anti-patterns. The "ThreatGet" tool is being further developed in collaboration with industry partners, private customers, and public projects. Since the tool is already commercially distributed in partnership with Lieber Lieber, the development is not entirely independent of external requirements. Nevertheless, the following two features have been identified and will be integrated into the development process. \\

The first future development focuses on the integration of threat dependencies during the analysis. Currently, individual threats can be mapped and analyzed with the analysis language. However, threat and vulnerability dependencies cannot yet be fully integrated into the analysis.  As described initially, attacks rarely consist of single actions that lead the attacker to the target. In most cases, several attack steps that build upon each other must be conducted in order to achieve the actual goal \shortcite{navarro_systematic_2018}. These individual attack steps together form a more complex attack path that describes the entire attack. In order to map such an attack path with the help of the analysis language, it is extended by the concept of pre-and post-conditions \shortcite{lallie_review_2020}. This concept allows the analyst to depict the requirements as well as the resulting effects of a threat in an intuitive way \shortcite{aksu_automated_2018}. For example, conditions such as the required proximity of the attacker, or required knowledge can be integrated into the analysis and refine the analysis result. \\
This concept has already been evaluated in the master thesis of Korbinian Christl, in cooperation with the University of Vienna.  \\

The second extension of the ThreatGet tool concerns the system-model used, i.e., the diagram used to represent the system under investigation in an abstracted form. As mentioned in \autoref{sec:related}, different diagram models and approaches can be used to represent the security-related and security-critical aspects of a system. The currently used "Advanced Data-Flow Diagram" (ADFD) provides an easy entry point to modeling such a system due to its simplicity and clear structure \shortcite{shostack_threat_2014}. However, we have found in several use-case studies that supporting additional diagram models would improve the efficacy of the tool. \\
Therefore, we evaluate the support of the so-called "Internal Block Diagram" according to the "SysML" standard. This approach has the advantage that this type of diagram is often used to plan and IT-system \shortcite{schaad_tam_2012}.  \\
Support for multiple diagram types reduces the additional effort required by the user to transfer their system to the ADFD schema. Adapting the system-model would not require any change to the analysis language. Nevertheless, the semantic analysis may need to be adapted to the data model of the diagram.

\clearpage
\bibliographystyle{apacite}
\bibliography{main.bib}

\clearpage

\end{document}